\documentclass{emulateapj}
\usepackage{float}
\usepackage[caption=false]{subfig}
\usepackage{graphicx}
\usepackage{amsmath}
\shorttitle{White light sources in the 2011 Feb 15 flare}
\shortauthors{Kerr \& Fletcher}

\begin{document}

%%%%%%%%%%%%%%%%%%%%%%%%%%%%%%%%%%%%%%%%%%%%%%%%%%%%%
%%%%%%%%%%%%%%%%%%%%     TITLE & ABSTRACT     %%%%%%%%%%%%%%%%%%%%
%%%%%%%%%%%%%%%%%%%%%%%%%%%%%%%%%%%%%%%%%%%%%%%%%%%%%

\title{Physical Properties of White-Light Sources in the 2011 Feb 15 Solar Flare}
\author{G.~S.~Kerr and L.~Fletcher}
\affil{SUPA, School of Physics and Astronomy, University of Glasgow, G12 8QQ, Scotland, U.K.}
\email{g.kerr.2@research.gla.ac.uk}

\keywords{Sun: chromosphere --- Sun: flares --- Sun: photosphere}

\begin{abstract}

White light flares (WLFs) are observational rarities, making them understudied events. However, optical emission is a significant contribution to flare energy budgets and the emission mechanisms responsible could have important implications for flare models. Using  \textsl{Hinode} SOT optical continuum data taken in broadband red, green and blue filters, we investigate white-light emission from the X2.2 flare SOL2011-02-15T01:56:00. We develop a technique to robustly identify enhanced flare pixels and, using a knowledge of the RGB filter transmissions, determined the source color temperature and effective temperature. We investigated two idealized models of WL emission - an optically thick photospheric source, and an optically thin chromospheric slab. Under the optically thick assumption, the color temperature and effective temperature of flare sources in sunspot umbra and penumbra were determined as a function of time and position. Values in the range of 5000-6000K were found, corresponding to a blackbody temperature increase of a few hundred kelvin. The power emitted in the optical was estimated at $\sim 10^{26}$~ergs s$^{-1}$. In some of the white-light sources the color and blackbody temperatures are the same within uncertainties, consistent with a blackbody emitter. In other regions this is not the case, suggesting that some other continuum emission process is contributing. An optically thin slab model producing hydrogen recombination radiation is also discussed as a potential source of WL emission; it requires temperatures in the range 5,500 - 25,000K, and total energies of $\sim 10^{27}$~ergs s$^{-1}$.

\end{abstract}

%%%%%%%%%%%%%%%%%%%%%%%%%%%%%%%%%%%%%%%%%%%%%%%%%%%%%
%%%%%%%%%%%%%%%%%%%%     INTRODUCTION      %%%%%%%%%%%%%%%%%%%%%
%%%%%%%%%%%%%%%%%%%%%%%%%%%%%%%%%%%%%%%%%%%%%%%%%%%%%

\section{Introduction}\label{sec:intro}

A solar flare with emission in the the visible continuum is classified as a white-light flare (WLF) \citep{1966SSRv....5..388S,1989SoPh..121..261N}. WLF enhancements in the optical continuum against the photospheric background are typically a few tens of percent \citep{2003A&A...409.1107M,2005ApJ...618..537C}, although it has been suggested that the WLF observed by \cite{1859MNRAS..20...13C} had an intensity double that of the quiet photosphere \citep{2003ApJ...595..483M}. The relative difficulty of observing a small, localized enhancement against the bright photospheric background led to the notion of WLFs as unusual occurrences \citep{1989SoPh..121..261N}. Though still difficult to observe, it is now clear that WLFs are not as rare as previously thought and are not just a `big flare' phenomenon.  WLFs have been observed down to \textsl{GOES} class C1.6 \citep{2006SoPh..234...79H} using \textsl{TRACE} data, and recently the superposed epoch analysis of \cite{2011A&A...530A..84K} suggests that most flares from \textsl{GOES} classes C to X exhibit continuum enhancement. The optical continuum can account for a significant proportion of the total energy budget in a flare \citep{2004GeoRL..3110802W,2006JGRA..11110S14W}, though with significant uncertainties particularly in small events \citep{2011A&A...530A..84K}. Thus, optical emission is key to understanding flare physics, and particularly flare energetics. Identification of the emission mechanisms could shed light on energy transport in solar flares in general.  
  
To be classified as a WLF, an event must exhibit enhancement in the optical \textsl{continuum}. Roughly speaking, WLFs have enhancements in the continuum at wavelengths $\lambda> 3600$\thinspace \AA\ \citep{1989SoPh..124..303M,1989SoPh..121..261N}. That is, WLFs are enhancements in the Paschen continuum ($\lambda > 3647$\thinspace \AA), with contributions from part of the Balmer continuum ($912$\thinspace \AA\ $ < \lambda < 3646$\thinspace \AA). The intensity contrast tends to be most pronounced in the blue - \cite{1966SSRv....5..388S} reported WLF ribbons as having a ``blueish white color'' - and in many cases declines towards longer wavelengths \citep{1966SSRv....5..388S,1983SoPh...85..285N}. Contrasts in the red part of the visible spectrum can be three times smaller \citep{1989SoPh..121..261N}.  In some WLFs the Balmer jump at $ \lambda = 3646$\thinspace \AA\  can be observed, indicating hydrogen recombination radiation \citep{aller_63, osterbrock_astro}. `Type I' WLFs exhibit a Balmer jump, while `Type II' do not  \citep{1986lasf.conf..483M}. Three spectra from different WLFs, shown in \cite{1984SoPh...92..217N} illustrate that not all WLF spectra have the same properties, with the 10 September 1974 WLF  having a Balmer jump \citep{1982SoPh...80..113H}, the 7 August 1972 WLF  having no such jump \citep{1974SoPh...38..499M}, and the 24 April 1981 WLF (Neidig 1983) having a very smoothed-out jump near, but not at, the Balmer jump location \citep{1983SoPh...85..285N}.  There are insufficient broadband optical spectra to conclude whether the Balmer jump is a common feature.

\cite{1995A&AS..110...99F}  investigated three WLFs with HXR emission, two of which were 
considered Type I, and one of which was Type II, and noted several differences between them (as well as noting that there are WLFs which present as a combination of the two Types). The most noteworthy was that peak emission in the continuum of Type I WLFs correlate well in time with the peak emission in hard X-ray (HXR) and microwave emissions from the flare, whereas Type II WLF emissions are not seen to coincide, with a discrepancy of several minutes (in this case during the impulsive phase, but also reported in the gradual phase). The TRACE/RHESSI  flares discussed by \cite{2007ApJ...656.1187F} showed a strong tendency for good spatial and temporal correlations between white-light and HXR enhancements, as is also the case with recent Hinode flares \citep{2010ApJ...715..651W,2011ApJ...739...96K}. The close association in these cases between HXR and optical sources and timing suggests strongly that the non-thermal electrons that generate flare footpoint HXR emission are also responsible for the WLF emission. 

There is no consensus on the height of WLF emission in the solar atmosphere. Different models involved the upper chromosphere, the temperature minimum region (TMR) and the photosphere  \citep[][among others]{1972SoPh...24..414H,1986A&A...156...73A,2007ASPC..368..423F, 2006ApJ...641.1210X}. There are two main mechanisms proposed for the generation of WLF emission: (i) chromospheric free-bound emission and  (ii) enhanced photospheric H$^-$ continuum emission. Mechanism (ii) should show spectral properties consistent with a blackbody; (i) need not show this. Free-bound emission occurs when hydrogen  which is over-ionized during the flare, by accelerated electrons or other means, recombines \citep{1972SoPh...24..414H,2010MmSAI..81..637H}. The Balmer and (very weak) Paschen jumps have been observed in WLFs \citep{1989SoPh..124..303M}. This continuum would be optically thin; a property which has been demonstrated in one flare by \cite{2010ApJ...722.1514P}. An enhancement in the photospheric continuum due to an increase in H$^-$ opacity following photospheric heating would provide an enhanced blackbody spectrum. However, a significant problem is how to heat this region. In the context of the standard electron beam model,  an electron energy on the order of $100$~keV is required to reach the lower chromosphere \citep{1986A&A...156...73A}, and around a few MeV in order to reach the $\tau_{5000} = 1$ level.  It has also been proposed that the photosphere is heated by radiation originating in the chromosphere or above (radiative backwarming). \cite{1986A&A...156...73A} suggested that both emission mechanisms play a role in different wavelength domains, and that radiative backwarming would produce modest heating of the photosphere and temperature minimum region, with Balmer continuum or UV photons responsible for an increase in the temperature of the TMR of a few hundred kelvin. So, heating and ionisation in the chromosphere leading to the hydrogen free-bound continua can also produce a photospheric radiation enhancement. It may be the case that a combination of these mechanisms produces WLF emission, with the strength of each component varying from flare to flare. \\

In this paper, we carry out an analysis of WLF emission, and interpret it in the context of  two idealized models of WL emission -  optically thick radiation of photospheric origins, and an optically thin slab emitting free-bound radiation. In Section~\ref{sec:obser} we describe the observations used and Section~\ref{sec:data_red} the data reduction. The processing necessary to detect and characterise the optical flare sources is explained in Section~\ref{sec:optical_flare_sources} and the energetics calculations in \ref{sec:energ}. Model calculations are presented in Sections~\ref{sec:bb_energy} and \ref{sec:free_bound_rad}, and we end with Discussion and Conclusions.

%%%%%%%%%%%%%%%%%%%%%%%%%%%%%%%%%%%%%%%%%%%%%%%%%%%%%
%%%%%%%%%%%%%%%%%%%     OBSERVATIONS     %%%%%%%%%%%%%%%%%%%%%%
%%%%%%%%%%%%%%%%%%%%%%%%%%%%%%%%%%%%%%%%%%%%%%%%%%%%%

\section{\textsc{Observations}}\label{sec:obser}

\begin{figure}[h]
\centering
\includegraphics[width = 0.45\textwidth]{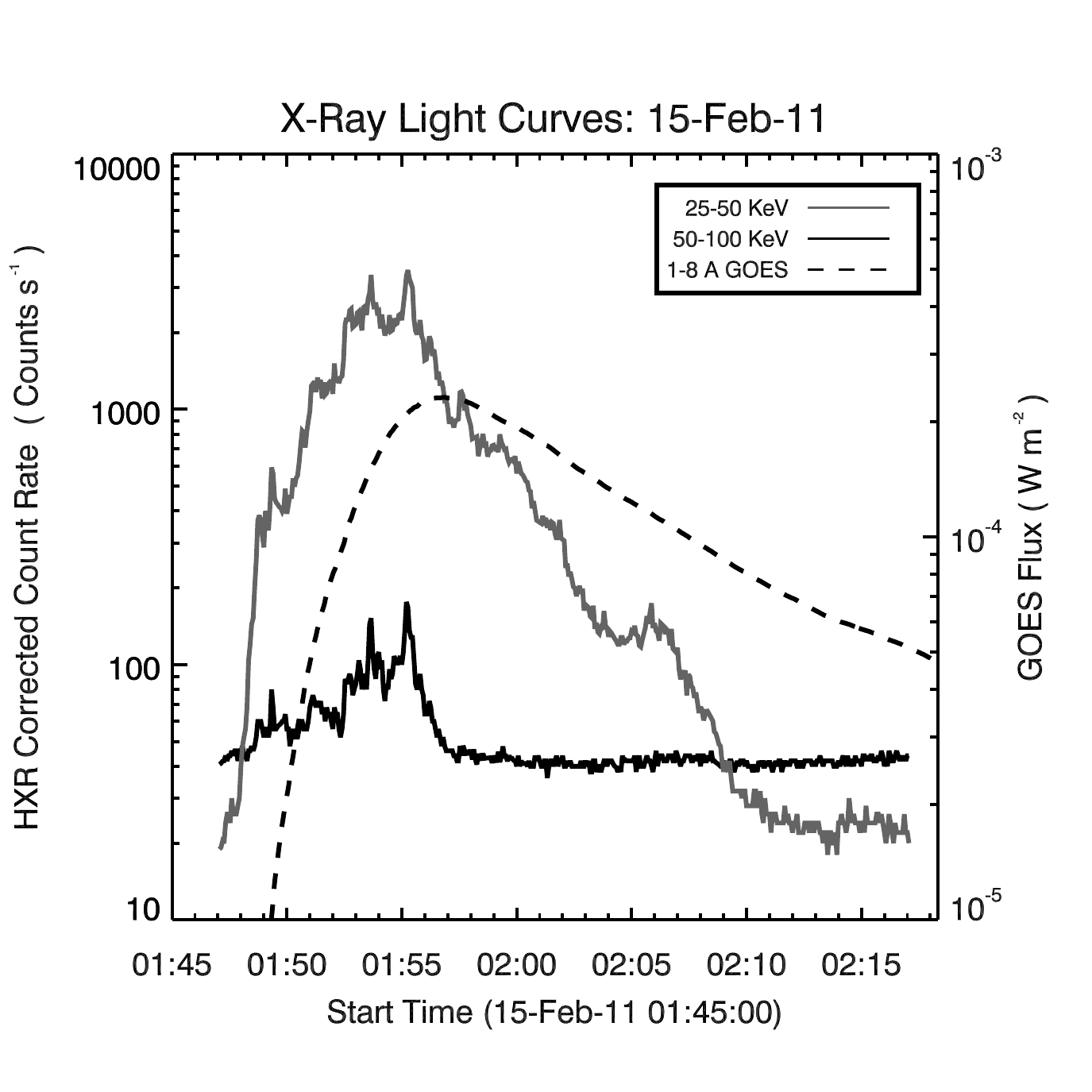}
\caption{\textsl{RHESSI and GOES lightcurves for the event}}
\label{fig:HXRlightcurves}
\end{figure}

The X2.2 flare SOL2011-02-15T01:56:00 occurred in NOAA active region 11158. It was the first X-flare of Cycle 24 and was well observed by various space telescopes. Several studies of the event have been performed using data from the Solar Dynamics Observatory \citep[SDO, ][]{2012SoPh..275....3P}, and the Ramaty High Energy Solar Spectroscopic Imager \citep[RHESSI, ][]{2002SoPh..210....3L}. Using measurements in the extreme ultraviolet (EUV) from the Extreme Ultraviolet Variability Experiment (EVE) on SDO, \cite{2012ApJ...748L..14M} found that the free-bound continuum at EUV wavelengths was well correlated with 25-50~keV RHESSI lightcurves, and suggested a chromospheric origin. \cite{2012ApJ...745L..17W} reported HXR footpoint sources either side of the polarity inversion line, which appear to follow the two flare ribbons. We focus on optical observations with the \textsl{Hinode} Solar Optical Telescope \citep[SOT, ][]{2008SoPh..249..197S}.  Figure~\ref{fig:HXRlightcurves} shows the GOES and RHESSI lightcurves for this flare.

The SOT Broadband Filtergram Imager (BFI)  observations between 01:50:21~UT and 01:59:42~UT captured the peak of the event. The pixel size of the BFI is 0.109~arcseconds, and observations were made in the red, green and blue continuum filters, at  $6684\pm 2$~\AA, $5550\pm 2$~\AA, and $4504\pm 2$~\AA~respectively, as well as in the Ca \textsc{ii} H line at $3968.5\pm 1.5$~\AA.~The BFI records sequentially at different wavelengths, and cycling through the three continuum wavelengths and emission lines gives a cadence per filter of around 19-21~s. 
Table~\ref{table:obser} displays a summary of the continuum data analyzed.  Level-0  data were downloaded from the \textsl{Hinode} Data Centre, and calibration was performed using the standard SOT routines in SolarSoft \citep{1998SoPh..182..497F}.

\begin{table}
\begin{center}
\begin{tabular}{l l l l}
\tableline \tableline
    & Red  & Green & Blue \\
\tableline \tableline   
\textbf{Wavelength} & ($6684\pm2$)~\AA\  & ($5550\pm2$)~\AA\ & ($4504\pm2$)~\AA\  \\
\textbf{Start Time} & 01:50:21~UT & 01:50:24~UT & 01:50:28~UT \\
\textbf{End Time} & {01:59:42~UT} & {01:59:45~UT} & {01:59:48~UT} \\
\textbf{Exposure Time} & {0.0512~s}  & {0.0768~s} & {0.06144~s} \\
\tableline
\end{tabular}
\caption{\textsl{Summary of 15-Feb-11 Level-0 Continuum}\label{table:obser}}
\end{center}
\end{table}

\begin{figure}
\centering
\includegraphics[width=0.5\textwidth]{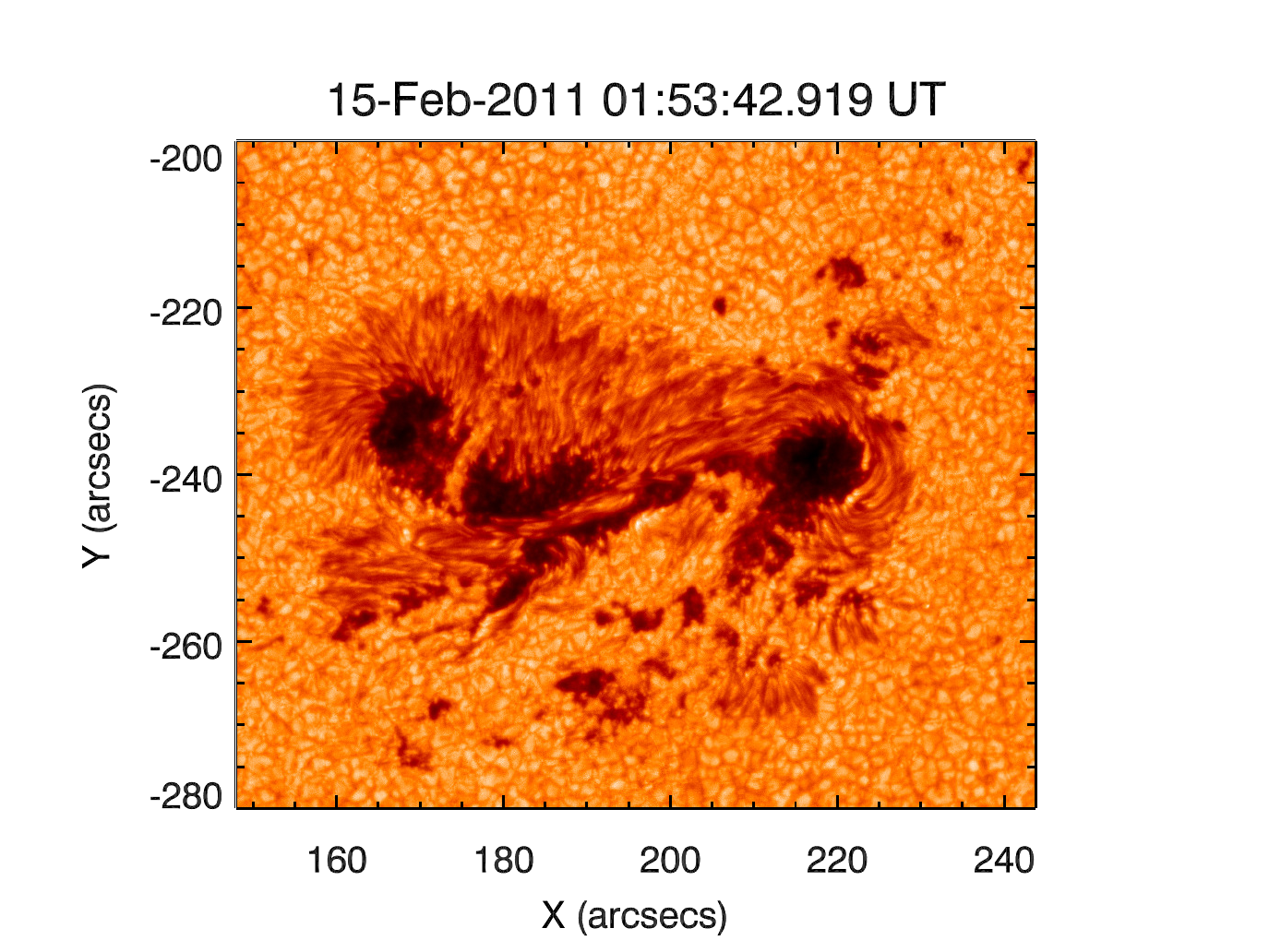}
\caption{\textsl{Calibrated {\emph{Hinode}}/SOT image made in the red continuum}}
\label{figure:red_image}
\end{figure}

The flare occurs in a rotating sunspot group with a complicated structure that includes penumbral fibrils with a twisted appearance, umbra only partly surrounded by penumbra, and light bridges, all seen in Figure~\ref{figure:red_image}. The penumbra, umbra and large scale features in the photospheric images have similar appearance across the three continuum wavebands. Transient features are likely to be emissions associated with flare footpoints or with granulation. The granulation causes noticeable variation to the background intensity which must be dealt with carefully when identifying the flare sources.

%%%%%%%%%%%%%%%%%%%%%%%%%%%%%%%%%%%%%%%%%%%%%%%%%%%%%
%%%%%%%%%%%%%%%%%%     DATA REDUCTION     %%%%%%%%%%%%%%%%%%%%%%%
%%%%%%%%%%%%%%%%%%%%%%%%%%%%%%%%%%%%%%%%%%%%%%%%%%%%%
\section{Data Reduction}\label{sec:data_red}

The level-0 data were pre-processed using the SolarSoftware (SSW) \textsl{Hinode} routine 
\texttt{fg$_{-}$prep}, which corrects for ADC offset, dark current, flat-field, and removes hot-pixels (e.g. cosmic rays) using a median filter. The counts (DN px$^{-1}$) were divided by the exposure time for each frame to give intensity measured in DN s$^{-1}$ px$^{-1}$. Typical exposure times are given in Table~\ref{table:obser} but vary slightly from frame to frame (on the order of 0.03\thinspace \%). Ca \textsc{ii} H images were also pre-processed using this routine.

For a quantitative analysis of the impulsive flare properties we require intensity in physical units, and the observed DN s$^{-1}$ px$^{-1}$ must be corrected for the telescope response function. Conversion factors from DN s$^{-1}$ to absolute specific intensity $I_{\lambda}$ in units of W~cm$^{-2}$ sr$^{-1}$ \AA$^{-1}$ were provided by Dr. T. Tarbell (Dr. T. Tarbell, \textsl{private communication} 2012). These calibrations by Dr. T. Tarbell were calculated as follows. The average solar disk spectrum given by the Brault \& Neckel 1987 Spectral Atlas available from Hamburg Observatory FTP site \citep{1999SoPh..184..421N}, was convolved with the wavelength-dependant instrument response in each SOT channel. Note the instrument response function from SSW is normalised, so the absolute response is then obtained by comparing the numerical value of the convolution with the observed quiet Sun disk centre intensity as measured on 14 February 2011. This calibration was performed averaging over many images of the quiet sun at disk centre, so the observed intensity (DN s$^{-1}$ px$^{-1}$) should correspond accurately to a solar spectral atlas (W~cm$^{-2}$ sr$^{-1}$ \AA$^{-1}$) at disk centre. 

This gives the power per unit area per unit solid angle over the bandpass of the channel, for each waveband. Intensities and conversion factors are given in Table~\ref{table:int_resp}.

\begin{table*}
	\begin{center}
	\begin{tabular}{c c c c}
		\tableline \tableline \\
			\multicolumn{1}{c}{{\textbf{Waveband}}} & 
			\multicolumn{1}{c}{{\textbf{Av. Disk Int.}}}&
			\multicolumn{1}{c}{{\textbf{Av. SOT Int.}}} &
			\multicolumn{1}{c}{{\textbf{Conversion Factor}}}  \\ 
			{}& (Wcm$^{-2}$sr$^{-1}$\AA$^{-1}$) & (DN~s$^{-1}$px$^{-1}$)  & {} \\		
		\\
		\tableline \tableline  
			{\textbf{Red}} &  {0.2742} & {36023.8} &  {7.6122 $\times$10$^{-6}$}\\
			{\textbf{Green}} & {0.3541} & 24236.6 & 1.4610 $\times$ 10$^{-5}$ \\
			{\textbf{Blue}} &{0.4316} & {22558.1}& {1.9133 $\times$ 10$^{-5}$}\\
		\tableline
	\end{tabular}
	\caption{\textsl{Disk intensities from the Brault \& Neckel spectral atlas, Average SOT intensities measured on 14-Feb-2011, and the resulting SOT conversion factors (Dr. Ted Tarbell, \textsl{private communication} 2012).}}
	\label{table:int_resp}
	\end{center}
\end{table*}

Image frames were co-aligned with the  01:52:43\thinspace UT (frame 10) using the small pore at [225, -215], just north of the leading spot. This was done in two steps. First, correction for the larger image jumps was done manually, by overlaying contours of each frame, in each waveband, (R, G, B and Ca \textsc{ii} H),  onto the base image and shifting until the contours lined up with the pore. The images were then cropped and each crudely coaligned waveband set was cross-correlated using the SSW routine \texttt{fg$_{-}$rigid$_{-}$align} to remove remaining small jitter. Visual inspection showed that coalignment was accurate to within 1-2 pixels.

%%%%%%%%%%%%%%%%%%%%%%%%%%%%%%%%%%%%%%%%%%%%%%%%%%%%%
%%%%%%%%%%%%%%%%%%     LOCATING WL PIXELS     %%%%%%%%%%%%%%%%%%%%%
%%%%%%%%%%%%%%%%%%%%%%%%%%%%%%%%%%%%%%%%%%%%%%%%%%%%%

\section{Optical Flare Sources}\label{sec:optical_flare_sources}
\subsection{\textsc{Locating White Light Flare Pixels}}\label{sec:findpixels}

\begin{figure*}
\begin{center}
\vbox{
\hbox{
\subfloat[{Difference image}]{\label{fig:diff1}\includegraphics[width = 0.43\textwidth]{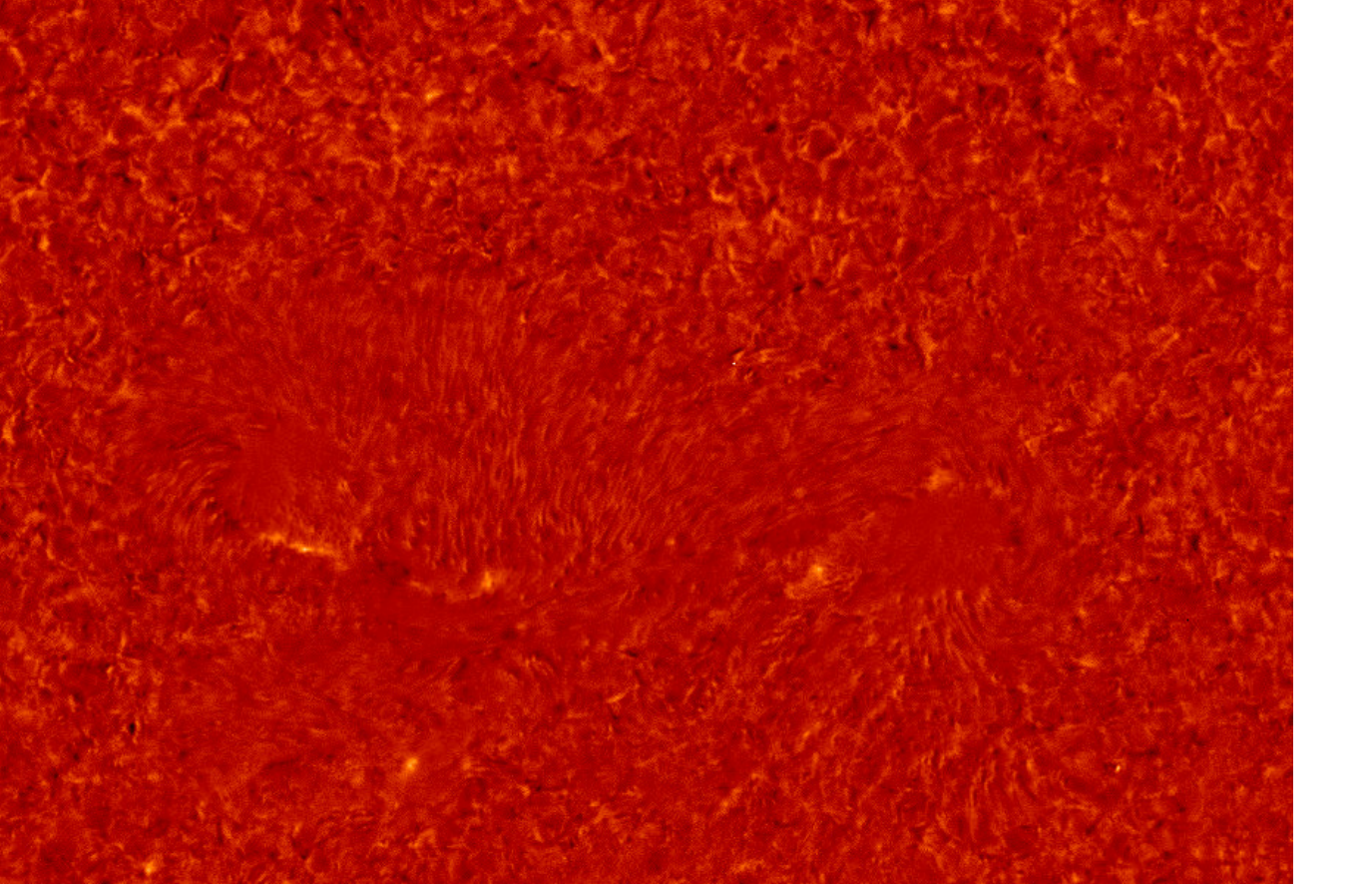}} 
\hspace{0.1\textwidth} 
\subfloat[{Filtered image}]{\label{fig:fil1}\includegraphics[width=0.43\textwidth]{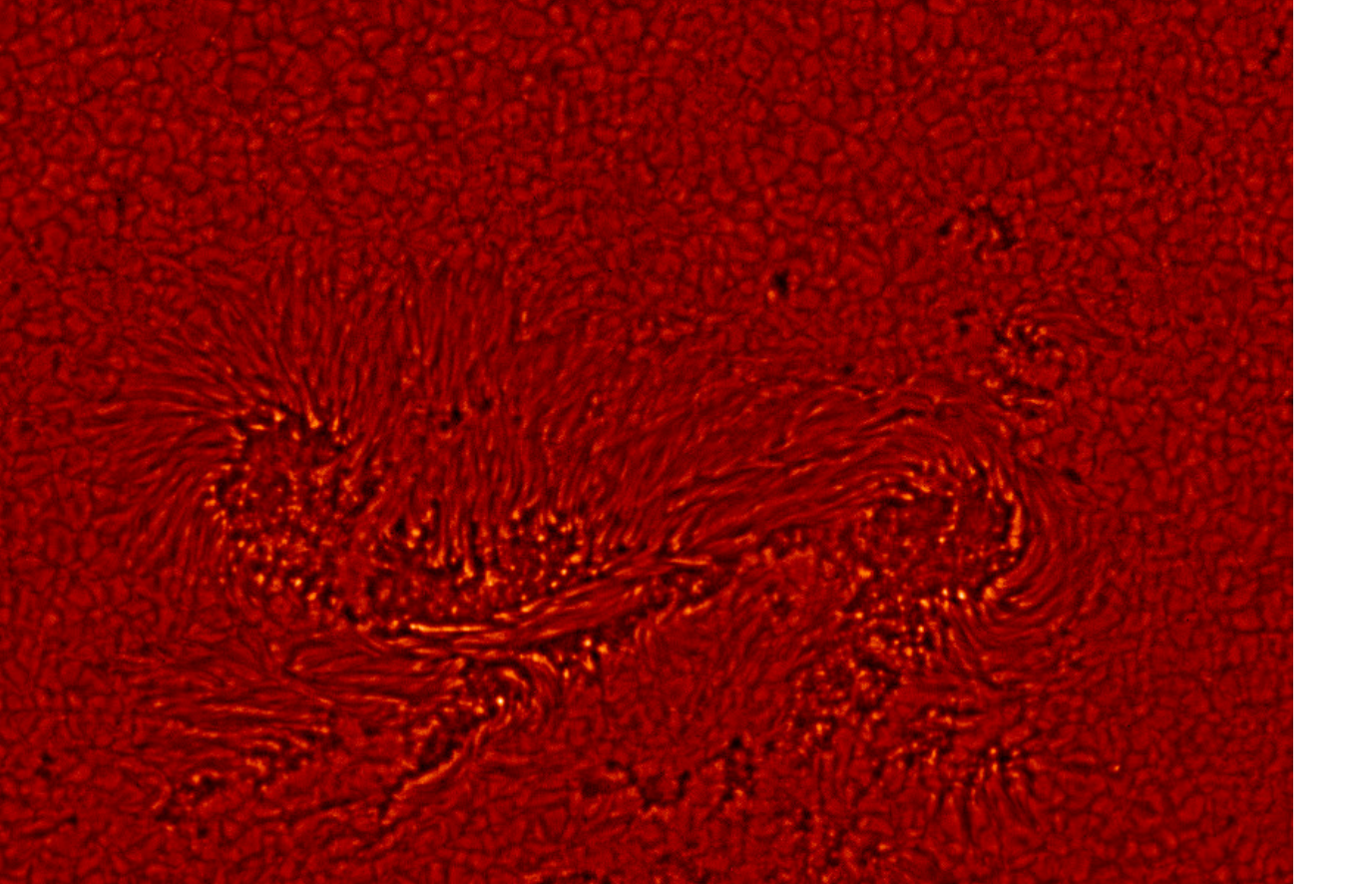}} 
}
\hbox{
\subfloat[{Difference of filtered image}]{\label{fig:dif2}\includegraphics[width=0.43\textwidth]{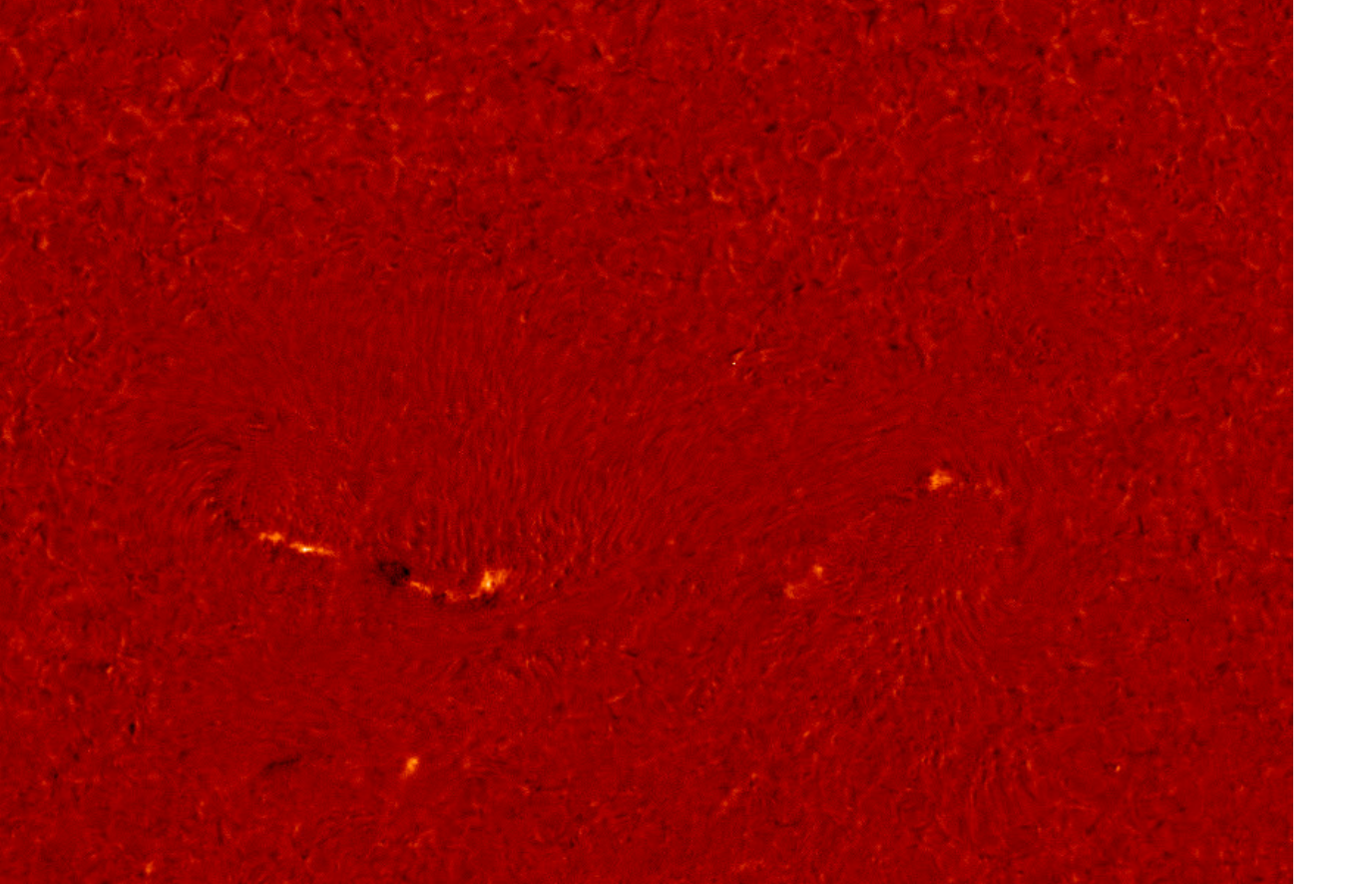}}
\hspace{0.1\textwidth} 
\subfloat[{Ca \textsc{ii} H emission}]{\label{fig:CaII}\includegraphics[width=0.43\textwidth]{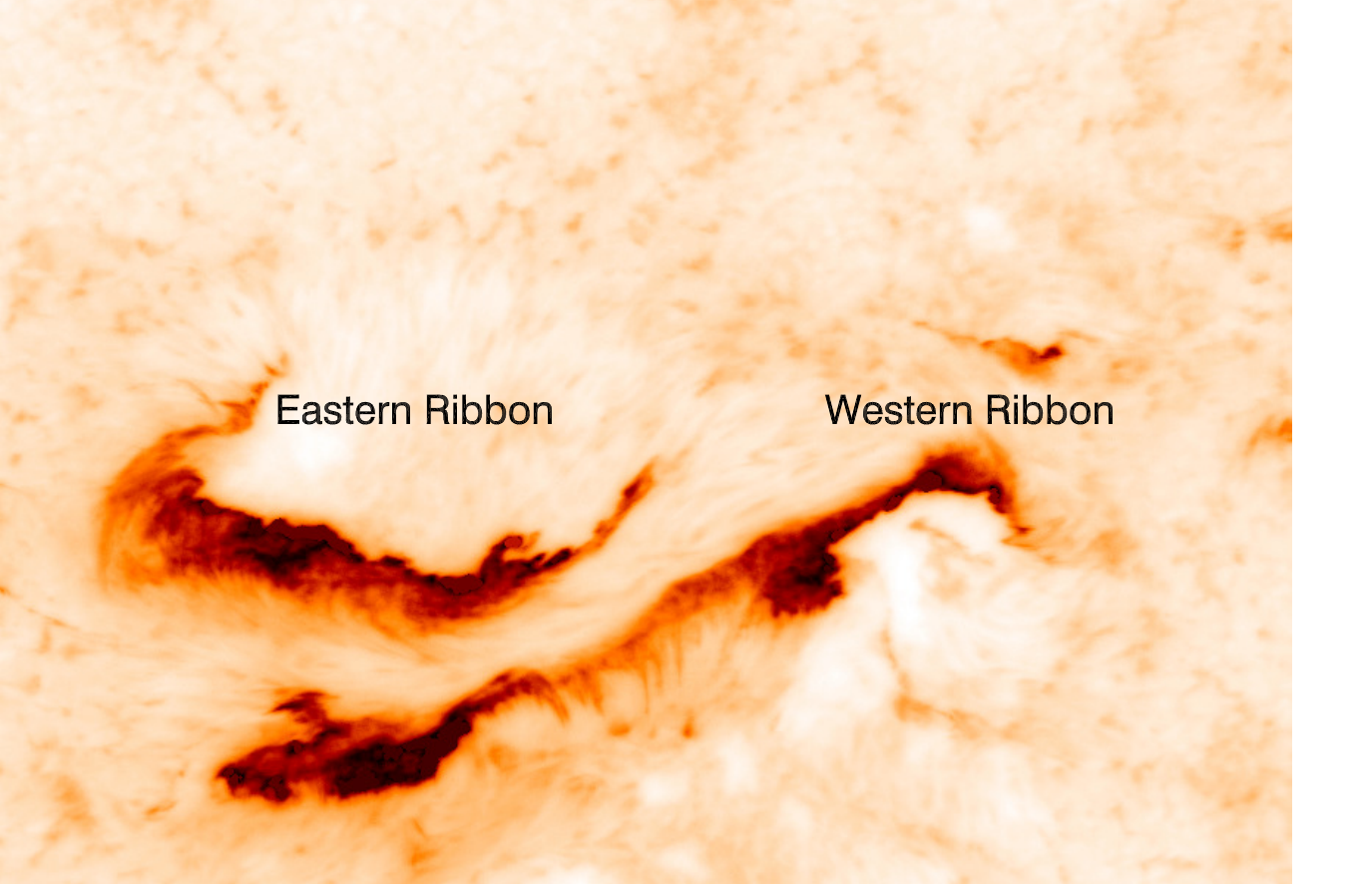}}
}
\hbox{
\subfloat[\small{Masked difference}]{\label{fig:dif3}\includegraphics[width=0.43\textwidth]{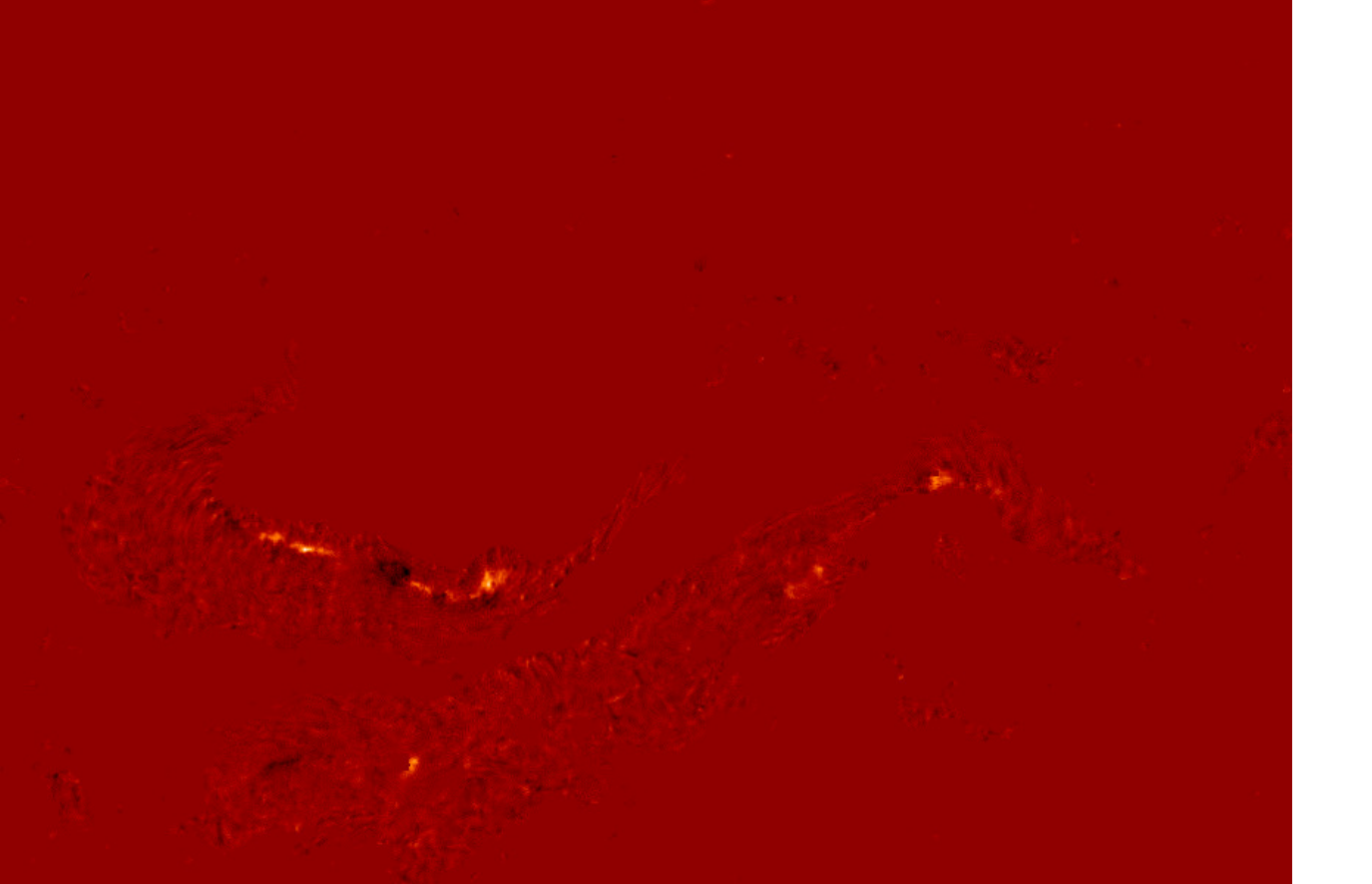}}
\hspace{0.1\textwidth} 
\subfloat[\small{Masked difference}]{\label{fig:dif4}\includegraphics[width=0.43\textwidth]{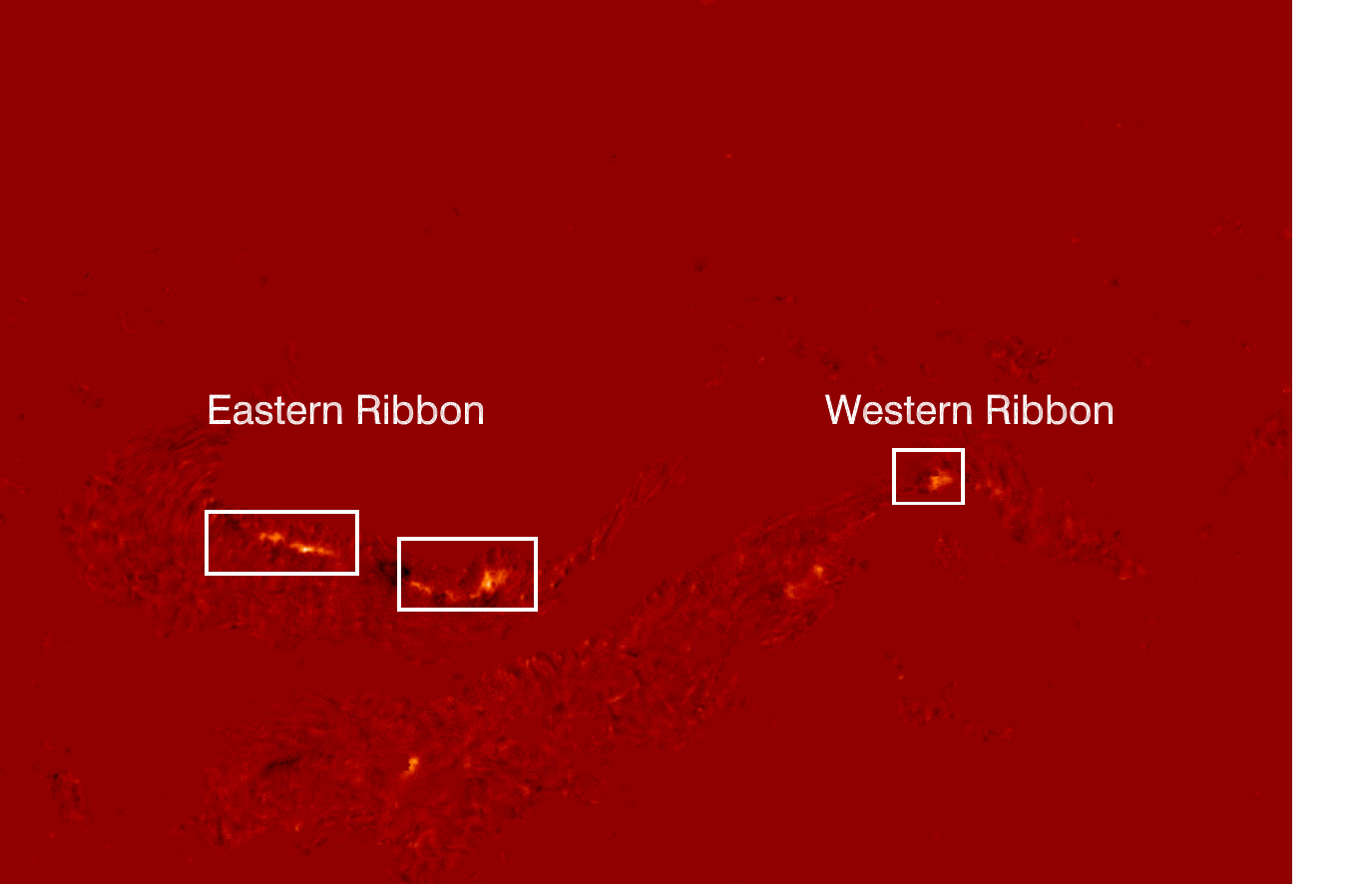}}
}
}
\caption{\textsl{Image processing steps for frame 10 in the red continuum: (a) A difference image, made by subtracting frame 8, illustrating the granulation noise. (b)  Log-unsharp mask filter applied to frame 10. (c) Difference image of the filtered frames 10 and 8. (d) The Ca \textsc{ii} H flare ribbons used as a mask. (e) The mask applied to image (c). (f) Examples of the pixel groups. Groups were then associated with either the eastern ribbon, or the western ribbon}}
\label{fig:processing}
\end{center}
\end{figure*}

Locating the newly flaring sources in each frame in the SOT continuum images is challenging due to the bright photospheric background, and difference imaging must be used in order to identify the transient, localized flaring sources. However, granulation in the SOT continuum means that base-differencing introduces significant and growing non-flaring signals to the difference images. This is somewhat removed by doing running differences, and we use (i-2) running differences as a compromise between reducing granulation noise and identifying clearly the flare changes. However, there is still some signal from the granulation, as can be seen in Figure~\ref{fig:processing}(a), so that applying a simple threshold to the difference images would still flag many non-flaring pixels.

In order to further reduce the appearance of granulation, before differencing the data were filtered using a log-unsharp filter. If $I$ is the logarithm of the original image, the log-unsharp filtered image is $I_{filter} = I - I_{smooth}$, where we used a 10-pixel boxcar smoothing. This filter enhances the edges of features, which helped to highlight the flaring sources as required. It also enhances the cell edges, but since granulation variability in the cell centers is larger than the edges, differencing these filtered images effectively suppressed  granulation noise. 

Figure~\ref{fig:processing}(b) shows the result of this process, and an (i-2) differenced image made from the filtered data in  Figure~\ref{fig:processing}(c) shows that both flare ribbons can be seen clearly, whereas the western flare ribbon was not so well defined in Figure~\ref{fig:processing}(a). A mask created from the Ca \textsc{ii} H emission seen in Figure~\ref{fig:processing}(d), in which flare ribbons can easily be identified using an intensity threshold, then further isolated flare brightenings. A masked, filtered difference image is shown in Figure~\ref{fig:processing}(e). The flare enhancements were most pronounced and most readily identified in the red filter images, so the flare pixel selection is made based on the red images. The reason that red sources were more identifiable in the difference images is likely due to granulation noise in the background. The red channel difference images were less noisy than green or blue channel difference images, though the reason for this is unknown. It is typically the blue wavelengths that show greater enhancement \citep{1989SoPh..121..261N}, and this indeed observed in the eastern ribbon for this flare, but not the western (c.f Section~\ref{sec:light_curves}).
For the subsequent temperature and intensity analyses, the same pixel locations are used in the (co-aligned) green and blue filters. 

Groups of WLF pixels were thus identified, and boxes drawn around the pixel groups (Figure~\ref{fig:processing}(f) illustrates three such pixel groups) and the mean and standard deviation of the processed (masked, filtered difference) images inside the each of the boxes were calculated. Pixels within each box with values greater than the local mean plus $n$ standard deviations were identified as WLF pixels, where $n$ was allowed to take values 1, 2, 3, or 4, to see the effects of different flare pixel detection criteria. Upon visual comparison to the difference images, it was decided that the 4-$\sigma$ detection threshold was too high; many flare pixels visible to the eye were not automatically identified. 
The WL emission appears as ribbon-like structures,  which are a subset of the Ca \textsc{ii} H ribbons in Figure~\ref{fig:processing}(d). The WL sources identified were `assigned' to a ribbon (East or West) depending upon their location in the image, in order to analyse the WL ribbons as separate structures, seen Figure~\ref{fig:processing}(f).

\begin{figure*}
\centering
\includegraphics[width = 0.8\textwidth]{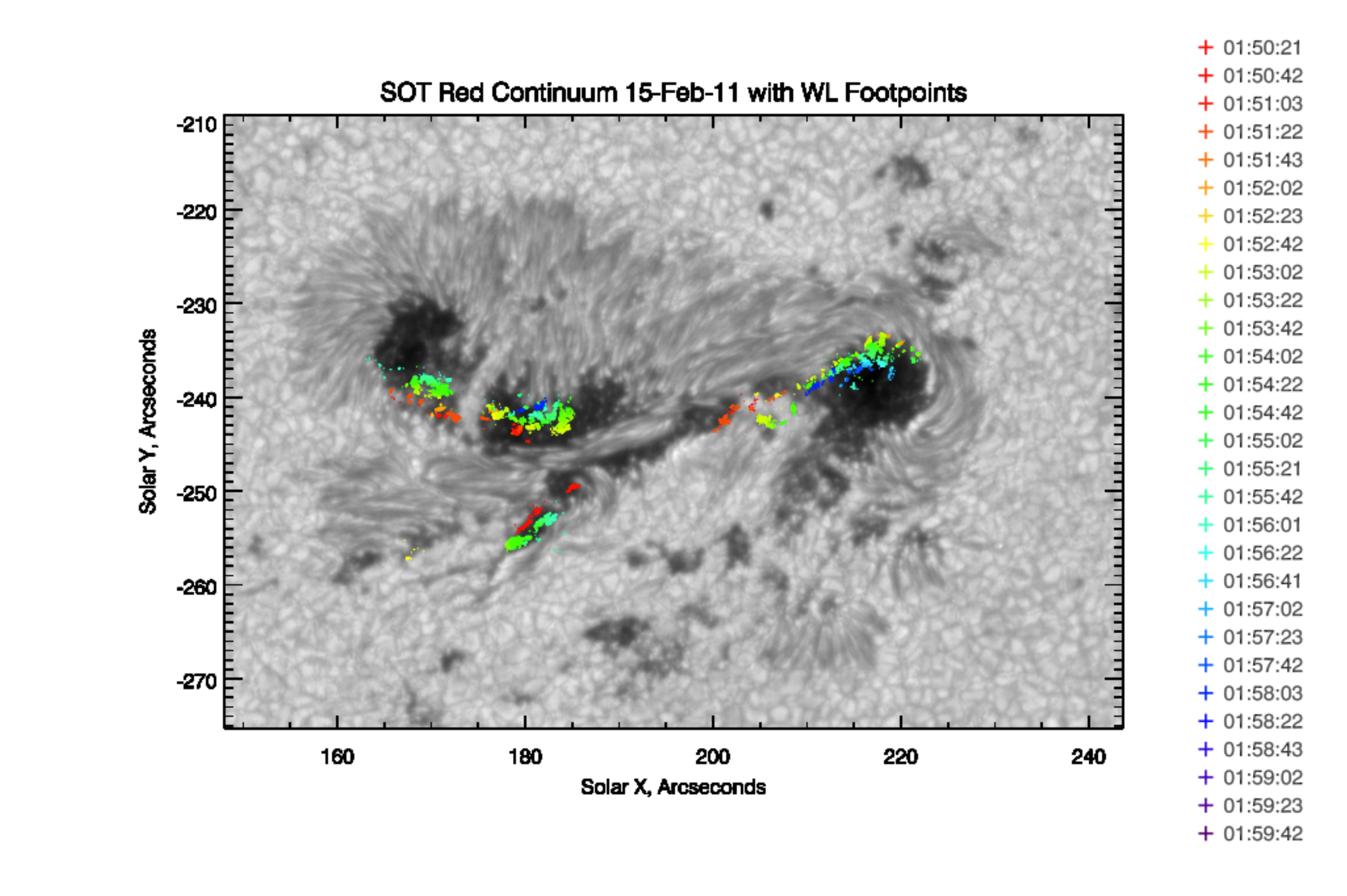}
\caption{\textsl{White light flare footpoints overlaid on a red continuum image.}}
\label{fig:fplocations}
\end{figure*}

Figure~\ref{fig:fplocations} shows the flare foopoints overlaid on the continuum data, where the color represents time. These pixels are identified using the criterion that their intensity is 2$\sigma$ above the mean in the processed data. There are three main ribbon sections. The eastern ribbon (ER) between $x = [160", 190"]$ and above $y = -246"$, is apparently split in two, being visible in the spot umbrae but not in the lightbridge area, though this may be simply due to low contrast in the lightbridge making detection difficult. The eastern ribbon moves to the north. The western ribbon (WR) is split in two, with one section (WR1) in the western sunspot, and the other (WR2) in a small portion of umbra, centred at $[180",-255"]$. The western ribbon moves to the south, into the umbra. WR2 was more difficult to observe, and less data was collected in this region (again, this could be due to contrast issues). 

Note that not every frame contained newly brightened WLF pixels and that after ~01:58UT no new WLF sources were identified.

Uncertainties on the intensity in any one frame are present due to uncertainties in identifying WLF pixels, $\sigma_{pixel}$. This can originate from both mis-identification of flare pixels locations and misalignment of frames. The size of $\sigma_{pixel}$ was estimated from the spread in intensities obtained by shifting the WLF pixel mask by 2 pixels, in each direction (x,y), resulting in 9 values for each shift (including the (0,0) shift in each case). The standard deviation of the intensities resulting from this shifts is $\sigma_{pixel}$. The values of  $\sigma_{pixel} $ were added in quadrature with photon counting error, $\sigma_{phot}$ providing the total uncertainty in WL intensity, $\sigma_{I}$.

%%%%%%%%%%%%%%%%%%%%%%%%%%%%%%%%%%%%%%%%%%%%%%%%%%%%
%%%%%%%%%%%%%%%%%%% RESULTS AND ANALYSIS   %%%%%%%%%%%%%%%%%%
%%%%%%%%%%%%%%%%%%%%%%%%%%%%%%%%%%%%%%%%%%%%%%%%%%%%

\subsection{\textsc{Optical Lightcurves}}\label{sec:light_curves}

As described in Section~\ref{sec:findpixels}, we have identified the \emph{newly brightened} pixels in each frame. For each set of new sources identified (e.g. those that first brighten in Frame \emph{i}), the temporal evolution of these WLF sources was measured, with the intensity defined by summing the intensity of the WLF pixels divided by the number of flaring pixels. In subsequent analysis we refer to the sources by the frame in which they initially brighten.

\begin{figure*}
\centering
\includegraphics[width = 1.\textwidth]{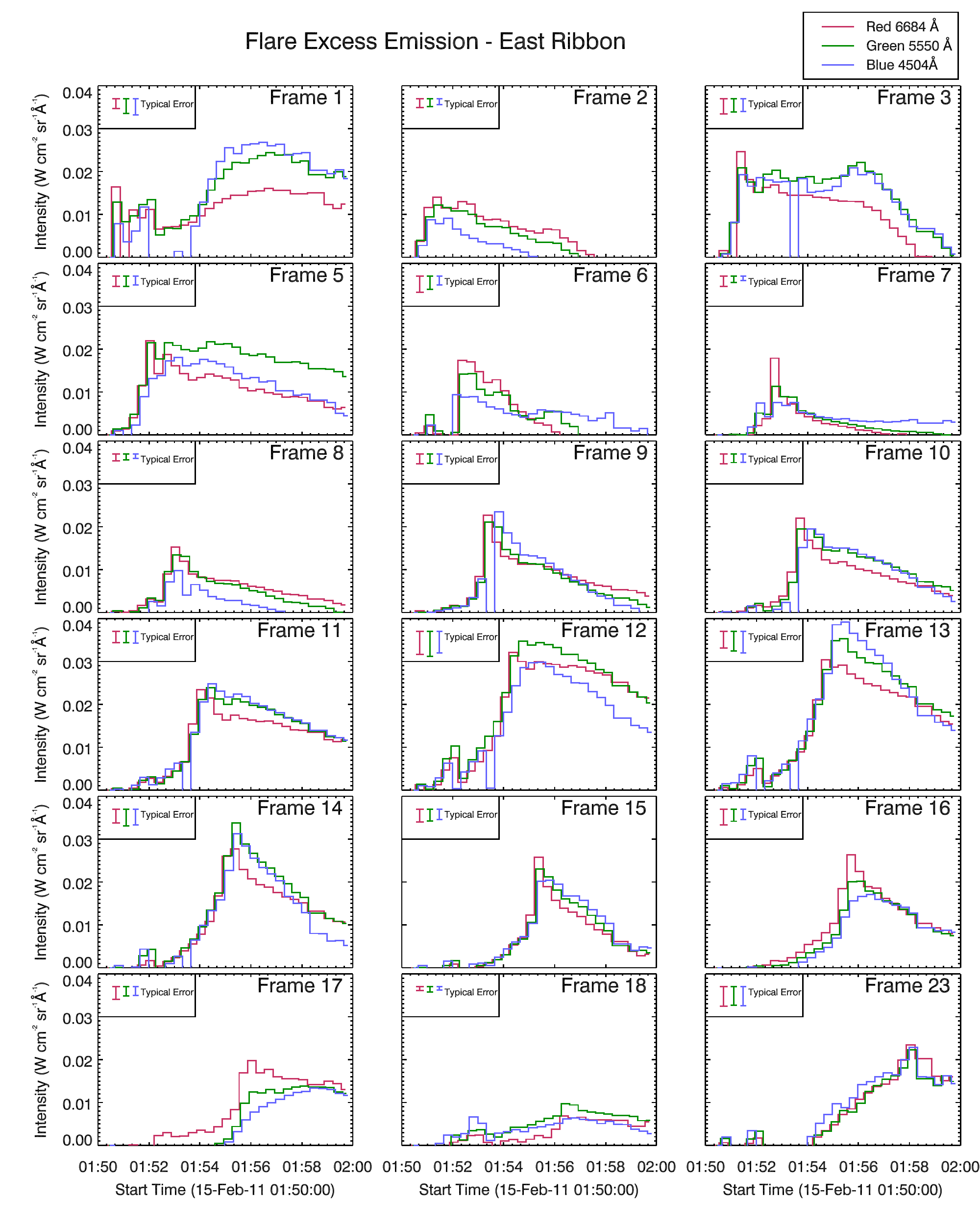}
\caption{\textsl{Lightcurves showing flaring emission from ER sources. The frame number refers to the frame in which sources were first identified. The typical error bar size in each frame is shown.}}
\label{fig:excess_east}
\end{figure*}

\begin{figure*}
\centering
\includegraphics[width = 1.\textwidth]{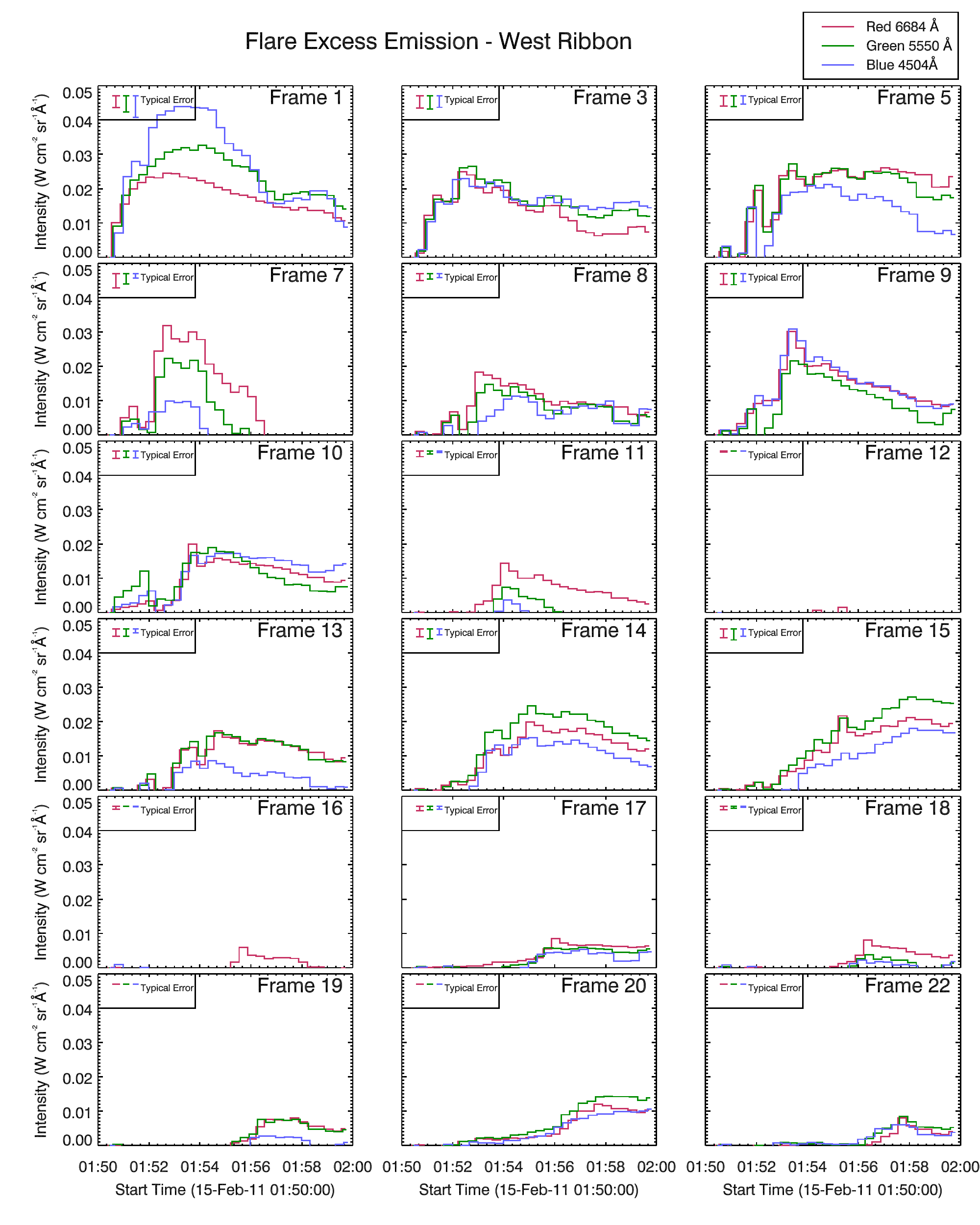}
\caption{\textsl{Lightcurves showing flaring emission from WR sources. The frame number refers to the frame in which sources were first identified. The typical error bar size in each frame is shown.}}
\label{fig:excess_west}
\end{figure*}

Flare excess lightcurves are shown in Figure~\ref{fig:excess_east} (east ribbon) and Figure~\ref{fig:excess_west} (western ribbon) - that is, the source background subtracted source intensity. Typical error bars are shown in the corner of each panel. Background subtraction for early frames was difficult because the flare had already began before SOT started making observations and post-flare observations were not available. However, since the flare progresses through the field of view it was possible to get an estimate of background intensity for sources that brighten in later frames. 

Some frames display ambiguous behaviour (particularly in the western ribbon, and in early frames for which a good background is not available), however around the peak of the HXR emission ($\approx$ frames 4 to 18) the excess lightcurves show clear flare signatures with a contrast of a few tens of percent observed. 

All three wavebands increase simultaneously within the cadence of each filter $\sim$20~s, though it is interesting to note that in some frames the red continuum passband seems to peak shortly before the green and blue (the reason for this is unknown and is presently under investigation). The rise is rapid,  with the impulsive peak lasting around 1-2 minutes, and is followed by a long decay. This is similar to the profiles associated with heating \citep{2007ASPC..368..365H,2010ApJ...725..319Q,2011SSRv..159...19F}.  Other authors \citep[e.g.][]{2003A&A...409.1107M,2008ApJ...688L.119J} noted similar timings for other WLFs. 

The regions remained enhanced for a significant period of time following the peak. The survey of \textsl{Yohkoh} WLFs by \cite{2003A&A...409.1107M} showed similar contrasts for energetic flares with footpoints in the penumbra or umbra, similar to our event. All three continuum wavelengths observed by the SOT are in the Paschen continuum (i.e. longwards of the Balmer jump), and previous observations suggest that lower contrasts occur there than wavelengths shortward of the Balmer jump  where 100\% contrasts have been seen \citep{1983SoPh...85..285N}. Due to the relatively low cadence of the data, it is also possible that the peaks have been smoothed, reducing the maximum intensity and the measured contrast. 

Type I WLFs are typically associated with the impulsive phase, whereas Type II WLFs can be associated with the gradual phase \citep{1995A&AS..110...99F}, and comparing with the HXR curves in Figure~\ref{fig:HXRlightcurves}, we see that all WLF enhancement occurs within the impulsive phase of the emission suggesting a Type 1 WLF.

Section~\ref{sec:energ}, which discusses the WL energy output from the flare, uses all of the identified WL emission. However, this WLF was difficult to detect, and was a reasonably weak WL emitter (evidenced by how difficult it was to detect), which could explain some of the features in the lightcurves, and so for the subsequent analysis of the physical properties of WLF emission (Section~\ref{sec:bb_energy} and Section~\ref{sec:free_bound_rad}) we present one set of flaring source from each ribbon that displays unambiguous behaviour. We select sources identified in Frame 11 from the eastern ribbon, and sources identified in Frame 9 from the western ribbon.
 
 \section{\textsc{Measured Optical Power}}\label{sec:energ}
Using the time history of each identified WL pixel we investigated the power radiated and the total energy released in each WL channel over the observation period.

The background-subtracted intensity, $I_{f,\lambda}$ was used to measure the WLF instantaneous power in each passband,  $P_{\lambda}$.  $I_{f,\lambda}$ is defined as the flare excess intensity emitted at $\lambda$ over a unit solid angle, per unit area $A$ across the width of the passband $\Delta \lambda$. Assuming isotropic radiation, the power emitted is then
			\begin{equation}
			\centering
			P_{\lambda} = \pi I_{f,\lambda} A \Delta \lambda
			\label{eq:powered}
			\end{equation}
 
In each frame the WLF pixels were identified and using Equation~\ref{eq:powered} the power of each pixel was calculated over the duration of the SOT observations. Integrating over time returned the energy produced by that pixel over the course of the observations. Summing the energy from each identified pixel gave the total energy emitted in each SOT passband over the almost 10 minutes of observations. Table~\ref{table:energy_full} shows energies on the order of 10$^{25}$~ergs in each passband. The passbands are around 4~\AA\  wide, so scaling up to a whole WL range of $\sim$2000~\AA\ the total energy in WL would be expected to be around three orders of magnitude greater (i.e on the order 10$^{28-29}$~ergs). If background subtraction is not carried out, an energy per SOT channel is typically 10$^{26-27}$~ergs over the 10 minutes observed.

No SOT observations beyond~01:59UT were available, but the lightcurves show a long decay phase that could extend for much longer. Additional energy may have been emitted as WL for an unknown period following the available observational time window, increasing the total energy. \cite{2012ApJ...748L..14M} report EUV emission in this flare with a decay that lasts for over an hour after the end of the SOT observations.

\section{\textsc{Optically Thick Modeling}}\label{sec:bb_energy}
\begin{table}
			\begin{center}
			\begin{tabular}{l l l l l l l}
			\tableline \tableline \\
				\multicolumn{1}{l}{} & 
				\multicolumn{3}{c}{} &
				\multicolumn{3}{c}{} \\
					{\textbf{Energy (ergs):}}& {\textbf{Red}} & {\textbf{Green}} & {\textbf{Blue}} \\
					\\
			\tableline \tableline
				{\textbf{\textbf{1$\sigma$ threshold}}} & 7.6$\times$10$^{25}$ & 7.3$\times$10$^{25}$ & 5.3$\times$10$^{25}$ \\ 
				{\textbf{\textbf{2$\sigma$ threshold}}} & 1.8$\times$10$^{25}$ & 1.7$\times$10$^{25}$ & 1.2$\times$10$^{25}$ \\ 
				{\textbf{\textbf{3$\sigma$ threshold}}} & 5.0$\times$10$^{24}$ & 3.5$\times$10$^{24}$ & 2.4$\times$10$^{24}$  \\
			\tableline
			\end{tabular}
			\caption{\textsl{Energy emitted in the SOT red, green and blue channels between 01:50 to 01:59. Rows correspond to values obtained setting different thresholds for identification of the flaring pixels.}}\label{table:energy_full}
			\end{center}
		\end{table}
 
We now return to the main topic of investigating the properties of the WL sources, using Frame 11 from ER, and Frame 9 from WR.

The first model considered was that of an optically thick blackbody source. A patch of local temperature enhancement would produce an increase to the WL radiation emitted by that patch from enhanced H$^-$ opacity. We do not address the source of the required heating. 

\subsection{\textsc{Source Temperature}}
The temperature of the WL sources was determined in two ways. The color temperature was obtained using filter ratios, and the RGB intensity values were also fitted to a blackbody curve, to determine the effective temperature $T_{\rm eff}$.  If WL emission was from a blackbody then all of these temperatures should be consistent within errors. In this analysis, the pre-flare background was not subtracted from the flare sources, consistent with the assumption of an optically thick source, the temperature of which increases during the flare.

\subsubsection{\textsc{Filter Ratio Method}}\label{sec:filter_ratio_method}
	
The Planck function
			\begin{equation}
			\centering
			B_{\lambda}(T) = \frac{2hc^{2}}{\lambda^{5}}\frac{1}{\exp \left(\frac{hc}{\lambda k_{b}T}\right) -1 }
			\end{equation}
was used to calculate model emission at 6684~\AA , 5550~\AA , and 4504~\AA . These are $B_{R_{6684}}$, $B_{G_{5550}}$, and $B_{B_{4504}}$, respectively, at temperatures in the range T~$\in$~[0, 20000]~K.  The observed intensities are the convolution of the Planck function with the filter passband function, however as the width of the passbands for each of the filters is the same and they are narrow and symmetric we can assume that the intensity does not vary significantly across the passbands (this assumption was tested against gaussian filters which are good approximations to the filter shapes, which resulted in differences of much less than 1\%). 

With three filters, two independent ratios can be calculated: 
			\begin{equation}
			\centering
			R_{BG}(T) = \frac{B_{B_{4504}}(T)}{B_{G_{5550}}(T)}, \qquad R_{BR} (T)= \frac{B_{B_{4504}}(T)}{B_{R_{6684}}(T)}
			\end{equation}
		
		\begin{figure}
			\centering
			\includegraphics[width=0.4\textwidth]{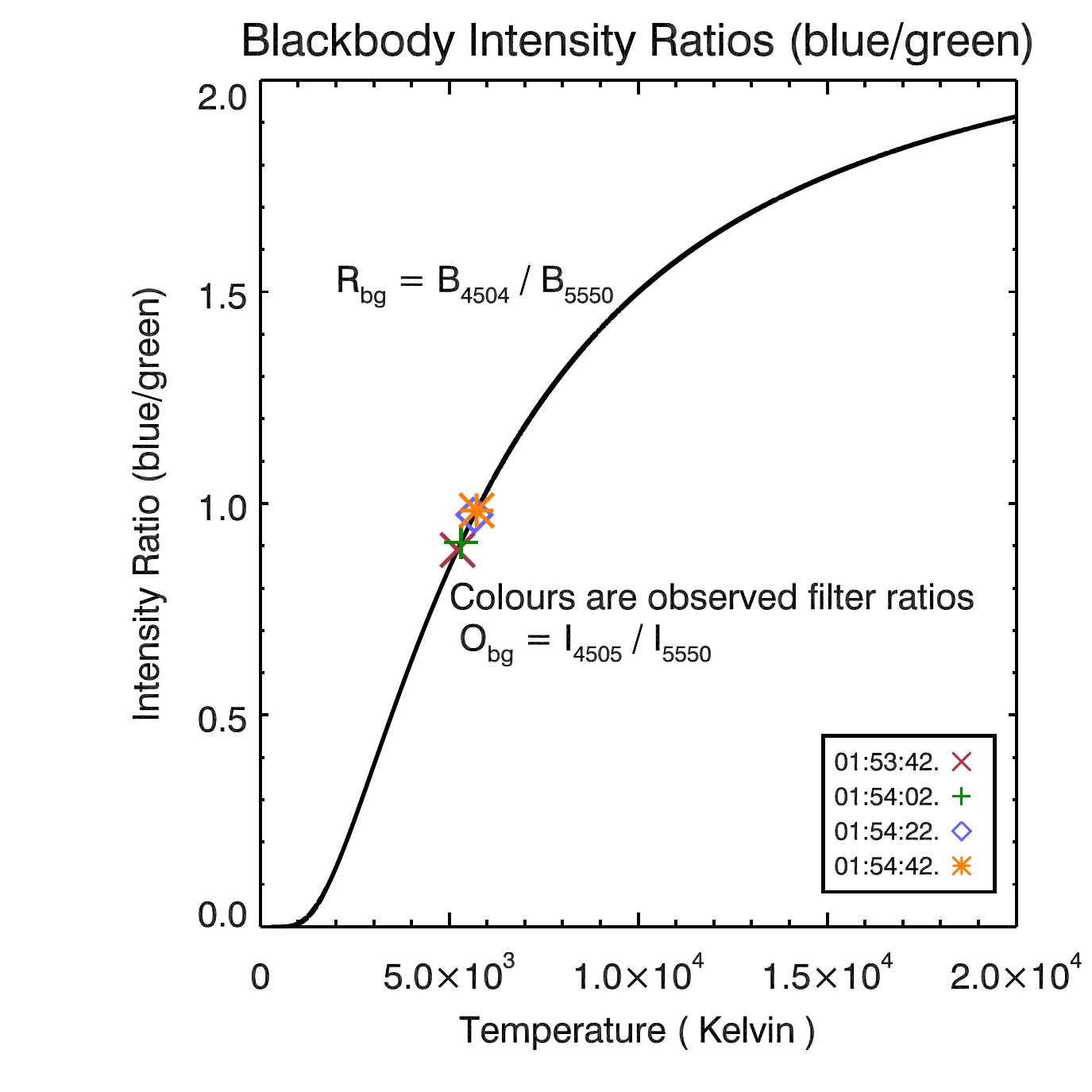}	
			\caption{\textsl{The variation of the SOT blue to green filter intensities as a function of blackbody temperature, illustrating the filter ratio method for determination of source temperature}}
		\label{fig:filterratios}
		\end{figure}	

The curve $R_{BG}$ is shown in Figure~\ref{fig:filterratios}, with the observed filter ratios at a number of time steps overlaid. As images in each filter were not simultaneous the ratios were made with the closest-in-time pairs, with RGB channels separated by $\approx$ (3-6)~s. The error, $\sigma_{time}$, in the measured ratio due to this lack of simultaneity was estimated by assuming that the intensity in each filter varied linearly between exposures (separated by $\approx$ (19-21)~s), providing an estimate of the maximum variation of intensity over the (3-6) seconds offset between exposures in different filters. The timing error and the error on intensity are summed in quadrature, and used to calculate the error on the filter ratios. Using the upper and lower estimates of the filter ratios, an error on the colour temperature was measured. These uncertainties are large, and are mostly due to systematic errors, which are the pixel detection and timing offset. It is hoped that a flare will be observed with WL sources that are easier to identify can help to reduce these systematic errors, in order to clarify the results.

Using these measured filter ratios, the temperature of the emitting region could be determined by comparing the results to the model filter ratios. This was done using a threshold for identifying the sources of 2$\sigma$ above the mean, but results were similar using 1- or 3-$\sigma$ thresholds.

\begin{figure}
\centering
\vbox{
\hbox{
\subfloat[East ribbon temperature]{\label{fig:LHtemp}\includegraphics[width=0.5\textwidth]{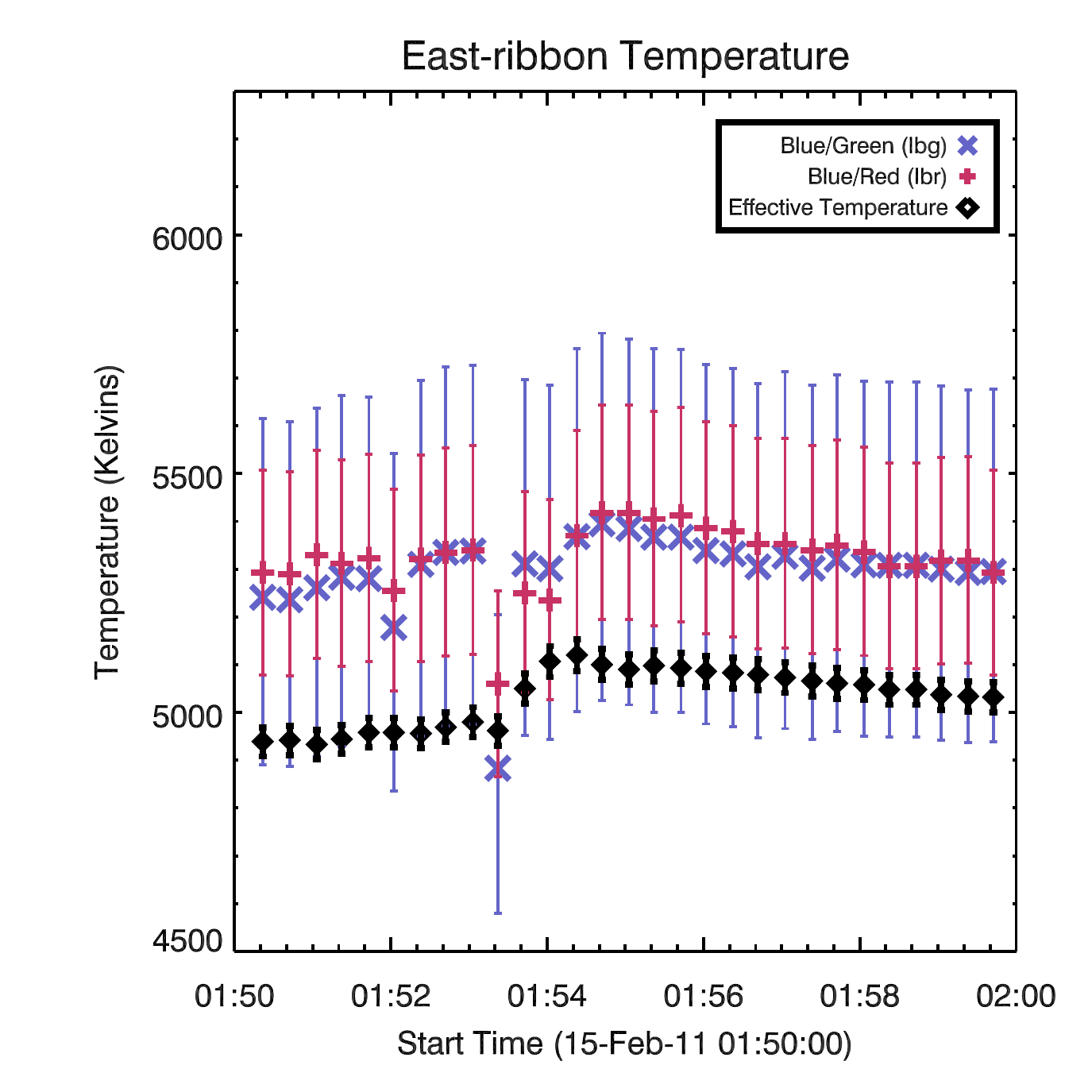} } 
}
\hbox{
\subfloat[West ribbon temperature]{\label{fig:RHtemp}\includegraphics[width=0.5\textwidth]{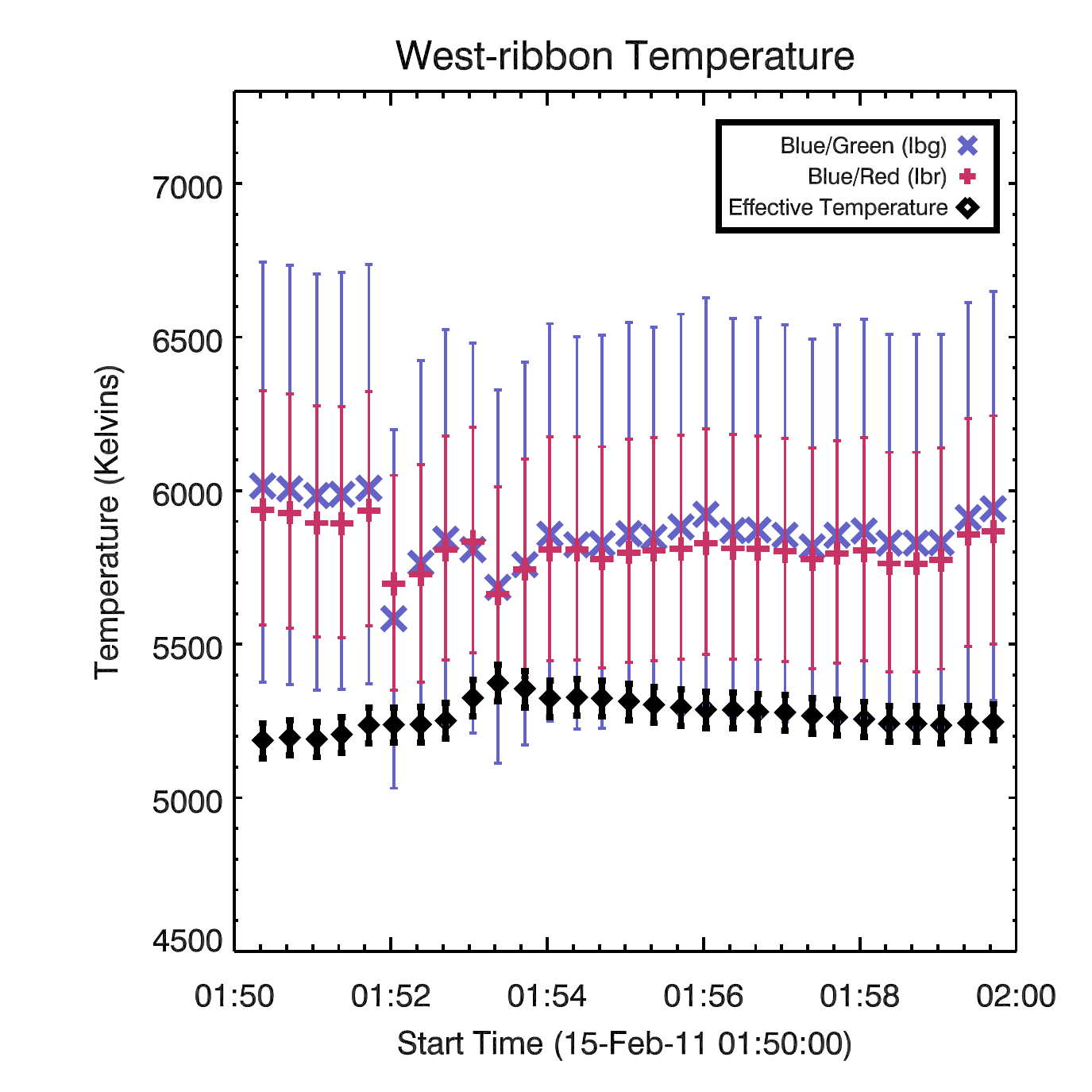}}
}
}
\caption{\textsl{Temperatures of WL sources from Frame 11 (ER) and Frame 9 (WR), determined using the filter ratio method, and the the blackbody fitting method. Coloured crosses show the colour temperatures, while black diamonds show the effective temperature.}}
\label{fig:temperatures}
\end{figure}

Figure~\ref{fig:temperatures}(a,b) shows the color temperature for each filter ratio measurement and the associated errors. Temperature increases of $\approx$200 Kelvin were observed.

The filter ratio results suggest that, to within the uncertainty in the data, the color temperatures were consistent in both the eastern and western ribbons, though the temperatures obtained from $R_{BR}$ were generally slightly higher than those from $R_{GB}$ in the eastern ribbon and vice versa in the western ribbon. This could correspond to the sources not being strictly blackbody but having temperature and opacity structure. For a discussion of the consistency with effective temperature see Section~\ref{sec:bbody_fit}.

\cite{2011A&A...530A..84K} performed a similar analysis with full Sun fluxes, using blue and green passbands, and found a WLF colour temperature larger than we observe with this flare ($\approx$9000~K compared with $\approx$5000-6000~K). The reason behind the difference between these temperature values is unknown, though we note that the flare analysed in our paper, though a very strong flare in HXR, was not a very strong WL emitter (as evidenced by difficulty of detecting the WL pixels). The WLFs identified by \cite{2011A&A...530A..84K} from full sun fluxes were likely to be have been very much stronger, being observable over the full Sun background. A stronger WL emitter would presumably be at a higher temperature (certainly if a blackbody source). Additionally, \cite{2013ApJ...776..123W} investigated a WLF occurring on 27 Janurary 2012, using Hinode/SOT, and measured a similar temperature to our results.

\subsubsection{\textsc{Blackbody Fitting}}\label{sec:bbody_fit}

Effective temperatures characterizing the SOT data were determined by fitting the Planck function, calculated for each of the SOT filters over the range T~$\in$~[0,20000]~K, and minimising the $\chi^{2}$ statistic: 

\begin{equation}
\centering
\chi^{2} =  \sum_{i=r,g,b} \frac{(I_{i} - B_{i})^{2}}{\sigma_{I,i}^{2}}
\end{equation}

Where $\sigma_{I}$ was the error on intensity. 

Figure~\ref{fig:temperatures}(a,b) shows the results of this (black diamonds) plotted with the filter ratio results for comparison. Errors on these temperatures are the blackbody temperatures for the upper and lower values of intensity. The eastern ribbon, for the most part, shows good consistency between the two-filter color temperatures and the effective temperature. While the magnitude of effective temperature is lower, the values are within the error of the colour temperatures. $\rm{T_{eff}}$ determined for the western ribbon shows the same overall behavior at late stages, but is not as large as the temperatures measured using the filter ratios. The western ribbon might not be emitting as a blackbody, or perhaps blackbody emission may be a weak component in combination with another source such as free-bound emission. Alternatively, the flare's position on the boundary of umbra/penumbra place it in a location of high temperature gradient (flare enhancement not withstanding) so that a single temperature is a poor characterization of the overall source. The reason for the difference between the behaviour of $T_c$ and $T_e$ in the western ribbon behaviour prior to flare onset is not known, but could be due to noise in the data, enhanced when dividing the intensities.

\subsection{\textsc{WL Power - Optically Thick Assumption}}\label{sec:thin_energy}

\begin{figure}
			\centering
			\includegraphics[width=0.4\textwidth]{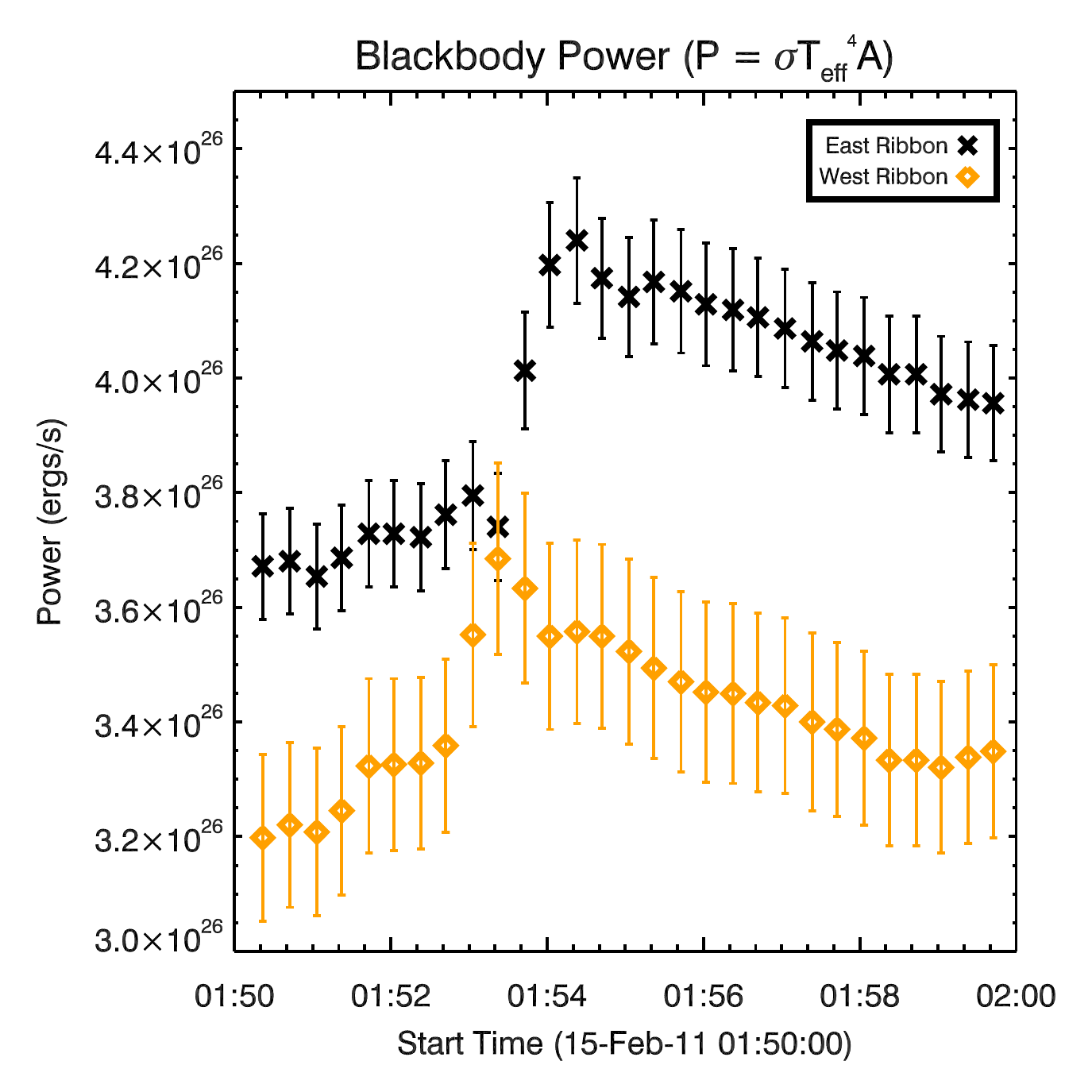}	
			\caption{\textsl{The power emitted by ER Frame 11 and WR Frame 9 based on calculated blackbody temperatures.}}
		\label{fig:blackbody_power}
		\end{figure}

Stefan-Boltzmann's law, $F = \sigma T_{\rm eff}^{4}$ was used to find the total integrated flux emitted by a blackbody at the temperatures $T_{\rm eff}$ obtained from the data. Using source areas thresholded at intensity levels n-$\sigma$ above the mean, the power $P = \sigma T_{\rm eff}^4 A$ was calculated for the three cases. This would be the energy emitted by WL sources per unit time during the impulsive phase. Figure~\ref{fig:blackbody_power} shows the blackbody power of WL sources (Frame 11 in eastern ribbon, and Frame 9 in western ribbon), based on temperatures derived in the previous section, which are typically $\approx 10^{26}~\rm{ergs~s^{-1}}$. 

\begin{figure}
			\centering
			\hbox{
				\subfloat[]{\label{fig:pl}\includegraphics[width=0.4\textwidth] {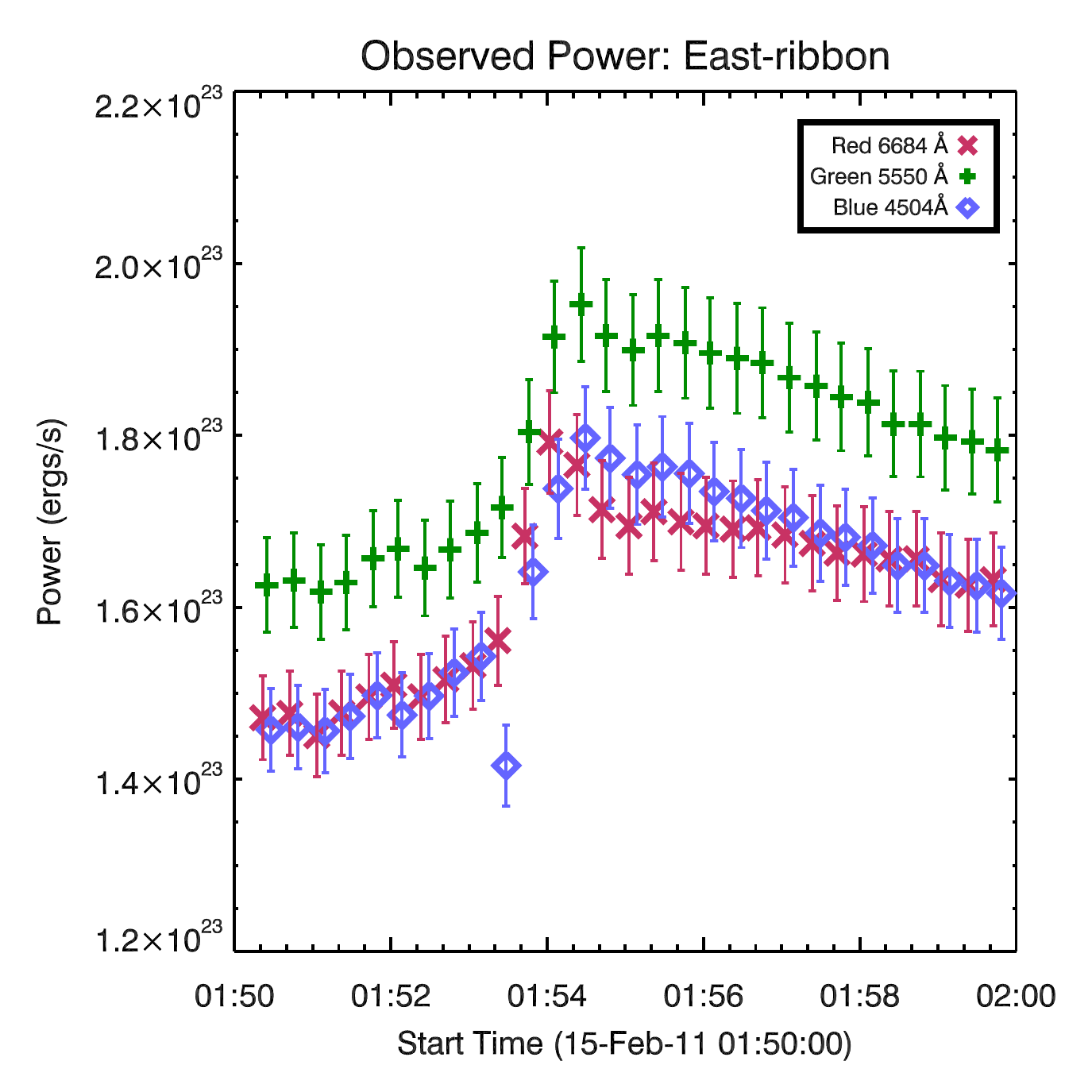}}
				}
			\hbox{
				\subfloat[]{\label{fig:pr}\includegraphics[width=0.4\textwidth]{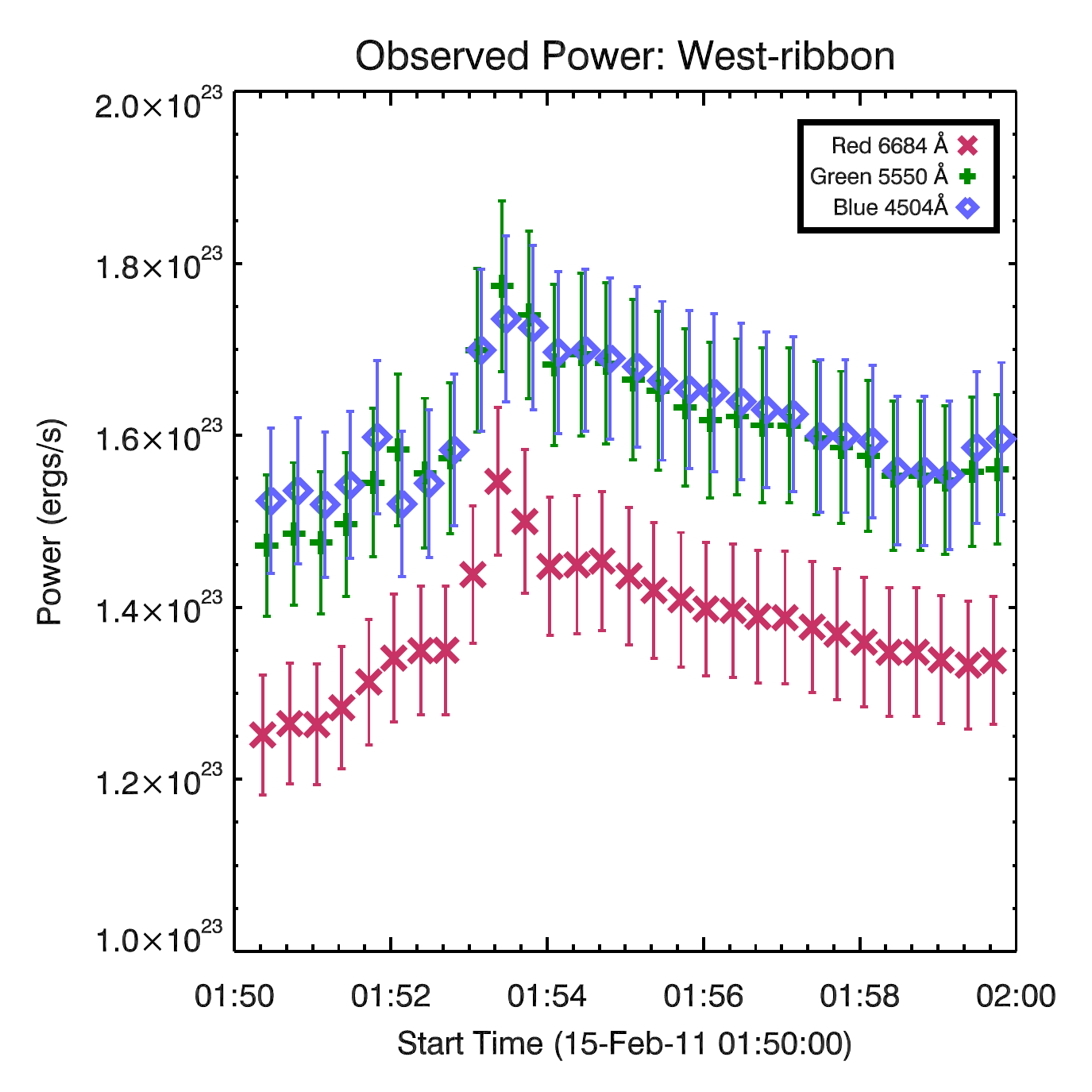}}  
				}
			\caption{\textsl{The instantaneous power emitted in WL calculated using 2$\sigma$ dataset.}}
		\label{fig:power}
		\end{figure}

We compare the {\emph{observed}} power of these WL sources with those calculated using the blackbody model with the effective temperatures. From the specific intensity per pixel, the total WLF instantaneous power in each passband,  $P_{\lambda}$, was measured, via Equation~\ref{eq:powered}, and shown in Figure~\ref{fig:power}(a,b) for the ER and WR.  The total error on intensity is described in Section~\ref{sec:findpixels}. $I_{\lambda}$, the full intensity measured (i.e. background + flare enhancements) was used in order to compare to the blackbody model. Power was typically on the order $\approx 10^{23}~\rm{ergs~s^{-1}}$, in each SOT filter. The power emitted over the whole WL spectrum would then be expected to be on the order of $10^{26}~\rm{ergs~s^{-1}}$. This is consistent with values obtained from the blackbody model.
	
\section{\textsc{Hydrogen Free-Bound Radiation}}\label{sec:free_bound_rad}
A simple chromospheric slab model was also investigated, which assumes that WL emission is produced from an optically thin source sitting above the photosphere and that no radiative backwarming of the photosphere was present. Subtracting the pre-flare intensities from the flare sources then determines the flare enhancements. This optically thin source was modelled simply as a slab with thickness L, uniform electron density n$_{e}$, and isothermal temperature T.

Following \cite{aller_63}, \cite{kowalski_thesis} and \cite{1982SoPh...80..113H} the hydrogen free-bound intensity, ${{I}_{\rm \lambda, fb}~\rm{ergs~s^{-1}cm^{-2} sr^{-1}} \AA^{-1}}$ is the product of the emission coefficient, $j_{\rm \lambda,fb}$ (ergs~s$^{-1}$~cm$^{-3}$~sr$^{-1}$~\AA$^{-1}$),  and the slab thickness, L (cm).
		
\begin{multline}\label{eq:Ifb}
%\begin{equation}%\label{eq:Ifb}
	I_{\rm \lambda,fb} = j_{\rm \lambda,fb}L = \left[\frac{6.48\times10^{-14}}{4\pi \lambda^{2}}\right]\left[\frac{n_{e}^{2}T^{-3/2}}{n^{3}}\right] \\ \exp\left\{\frac{1.58\times10^{5}}{n^{2}T} - 				\frac{1.44\times10^{8}}{\lambda T}\right\} L
%\end{equation}
\end{multline}\\

%\begin{align}
%	\begin{split}\label{eq:Ifb}
%I_{\rm \lambda,fb}  =\, &   j_{\rm \lambda,fb}L & \left[\frac{6.48\times10^{-14}}{4\pi \lambda^{2}}\right]\left[\frac{n_{e}^{2}T^{-3/2}}{n^{3}}\right] \\      
%                                   &\qquad \exp\left\{\frac{1.58\times10^{5}}{n^{2}T} - \frac{1.44\times10^{8}}{\lambda T}\right\} L
%  	\end{split}
%  \end{align}
%  

In Equation~\ref{eq:Ifb}, $\lambda$ is the wavelength (\AA), n$_{e}$ is the density (cm$^{-3}$), T is the temperature (K), and $n$ is the principal quantum number of the energy level in the hydrogen atom to which the electron recombines ($n$ = 3 in our case). 
		
This expression assumes ionisation equilibrium, a Maxwellian velocity distribution, that $\rm{n}_e = \rm{n}_i$, and the Gaunt factor is $\sim 1$. These assumptions should be kept in mind when interpreting results.\\

\subsection{\textsc{Parameter Constraints}}\label{sec:param_constraint}
There are three unknown parameters in Equation~\ref{eq:Ifb} - the temperature, density and slab thickness. Our assumption is that I$_{\lambda,fb}$  is the background-subtracted  intensity, which is known for  three SOT channels.  

\begin{figure}
\centering
\includegraphics[width=0.5\textwidth]{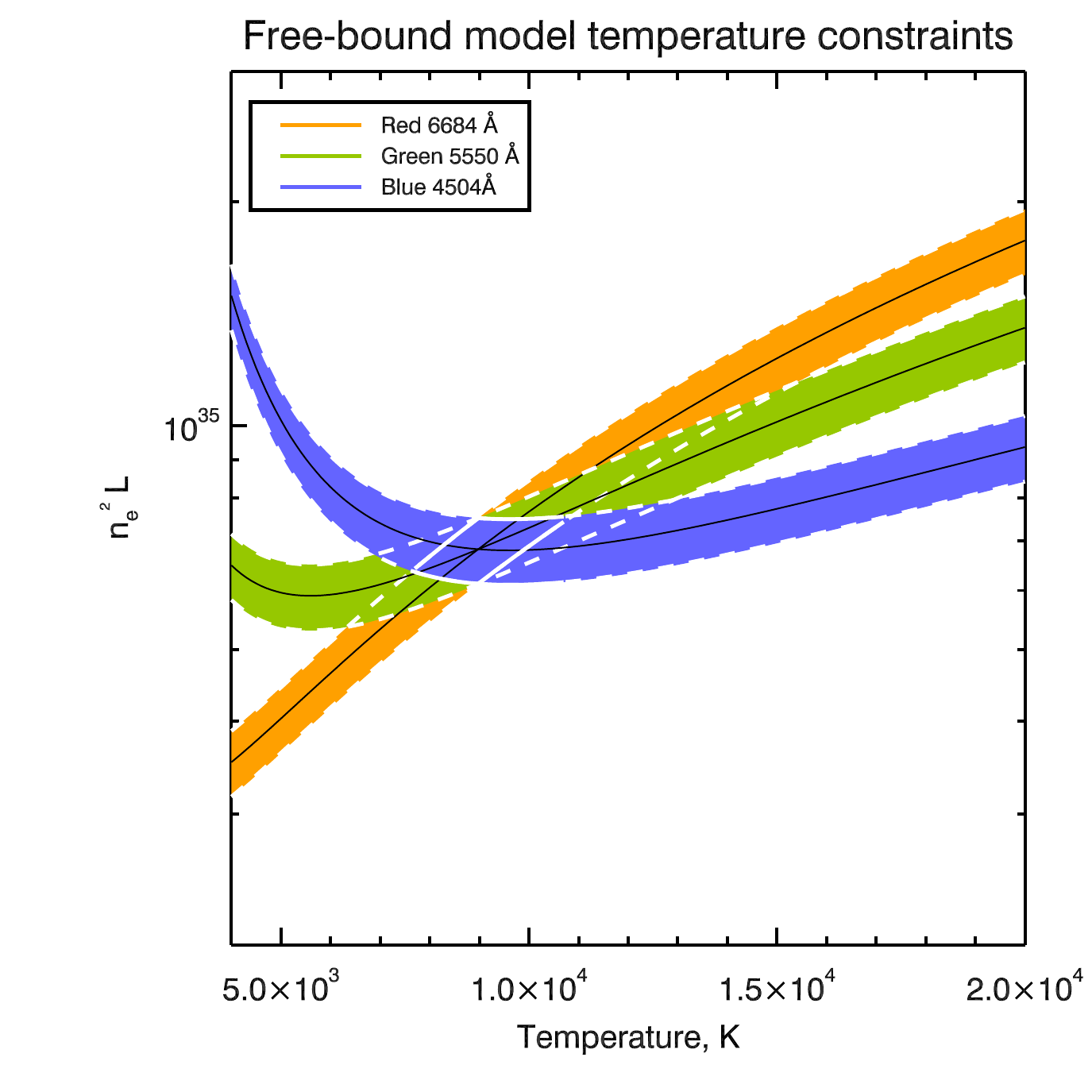}
\caption{\textsl{Example of temperature constraints in the simple free-bound slab model. Area of overlap highlights the range of possible values for n$_{e}^{2}$L, and is bounded by the upper and lower estimate of temperature.}}
\label{fig:val_err_left}
\end{figure}

Equation~\ref{eq:Ifb} is rearranged to Equation~\ref{eq:nel}: 

\begin{equation}\label{eq:nel}
n_{e}^2L = \frac{4\pi \times10^{14} \lambda^{2}}{6.48~\exp\left\{\frac{1.58\times10^{5}}{n^{2}T} - \frac{1.44\times10^{8}}{\lambda T}\right\}} n^{3}T^{3/2}I_{\rm \lambda,fb}
\end{equation}

and, for each of the observed $I_{\rm \lambda,fb}$, values of $n_{e}^2$L were calculated for a range of temperatures T $\in[4000, ~3\times10^{4} ]$~K. These will be referred to as n$\rm_{e}^{2}$L$\rvert_{\rm \lambda,obs}$ (the errors on the red, green and blue intensity measurements were as described in previous sections.) Examples of these values are plotted in Figure~\ref{fig:val_err_left}. An overlapping section of the curves corresponding to the red, green and blue filters indicates a common region of $n_e^2 L$ and $T$ constrained by all three filters. In the example shown, at temperatures above $2\times 10^4$~K, there are no values of $n_e^2 L$ and $T$ for which the curves overlap. For all frames, temperatures were typically in the range $\approx$ 5500-20,000~K, but could go as high as $\approx$ 25,000~K and low as $\approx$ 4500~K. 

The temperatures bounding this overlap region were recorded, as was the temperature that had the closest overlap (this point of overlap was not always as close as in Figure~\ref{fig:val_err_left}). It should be noted that this model produced overlap regions when the intensity excess was suitably large, usually around the flare onset. Prior to the first detection the flare in each frame, this method was not as accurate and sometimes no overlap regions were present. This could indicate that the WL emission at certain stages was not well described by this free-bound model. 

\begin{figure}
\centering
\includegraphics[width=0.5\textwidth]{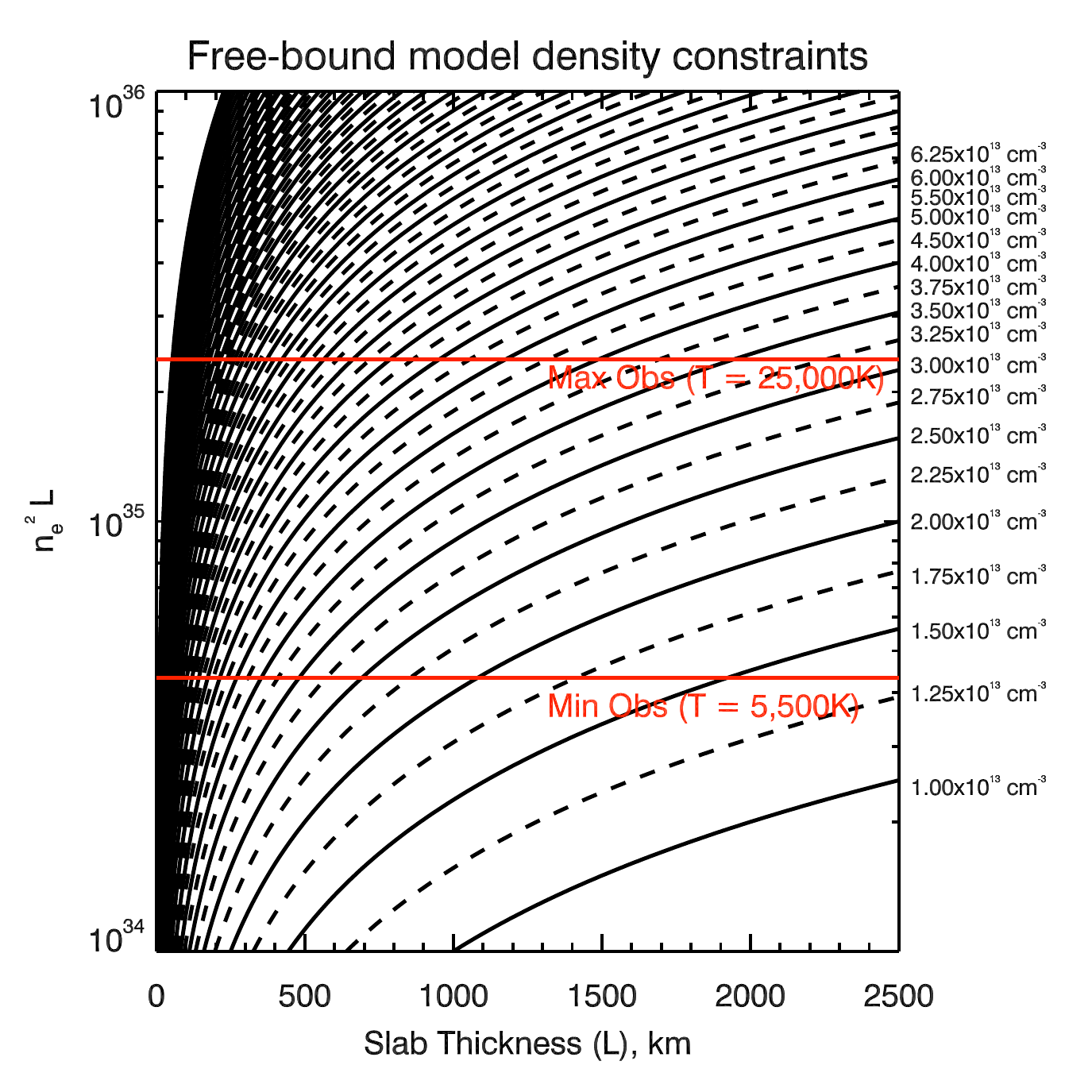}
\caption{\textsl{Range of possible values of n$_{e}^{2}$L$\rvert_{model}$ as a function of slab thickness. Curves are labelled according to the density used in the calculation. Horizontal lines show the  maximum and minimum observed values for $n_{e}^{2}$L$\rvert_{\lambda,obs}$ from a sensible temperature range based on Figure~\ref{fig:val_err_left}. }}
\label{fig:dens_constraint}
\end{figure}

The values of temperature bounding this overlap are the upper and lower limits permitted by both the model and observed intensity. Plotting values of $\rm n_{e}^{2}$L$\rvert_{\rm model}$ (calculated from a range of densities and slab thicknesses) against slab thickness, and overlaying the minimum and maximum value of n$_{e}^{2}$L$\rvert_{\rm \lambda,obs}$ for reasonable temperatures in Figure~\ref{fig:dens_constraint} indicates the range of possible values of density and slab thickness. 

Figure~\ref{fig:dens_constraint}, shows curves of $\rm n_{e}^{2}$L$\rvert_{\rm model}$ plotted for a range of $n_e$ values, and shows that for temperatures allowed by the model, $n_{e}$$\approx$ 10$^{13-14}$~cm$^{-3}$, 
depending on slab thickness. It is likely that the slab is in the thinner range of values plotted, suggesting a lower limit of around 10$^{13}$~cm$^{-3}$, for a slab a few hundred km thick, and an upper limit of around 10$^{14}$~cm$^{-3}$  for a thickness 100~km or less. According to models, such electron densities are found only low in the atmosphere in non-flaring chromospheres. However, they could be generated by complete ionisation of the middle chromosphere. Electron densities of $10^{13}\rm{cm}^{-3}$ have previously been found by \cite{1981ApJ...248L..39C}, using ions with a characteristic formation temperature of $60,000 - 100,000$K, and the inferred higher density and lower temperature we find is not inconsistent with this.

\subsection{\textsc{WL Power - Optically Thin Assumption}}\label{sec:thin_energ}
\begin{figure}
\centering
\includegraphics[width=0.5\textwidth]{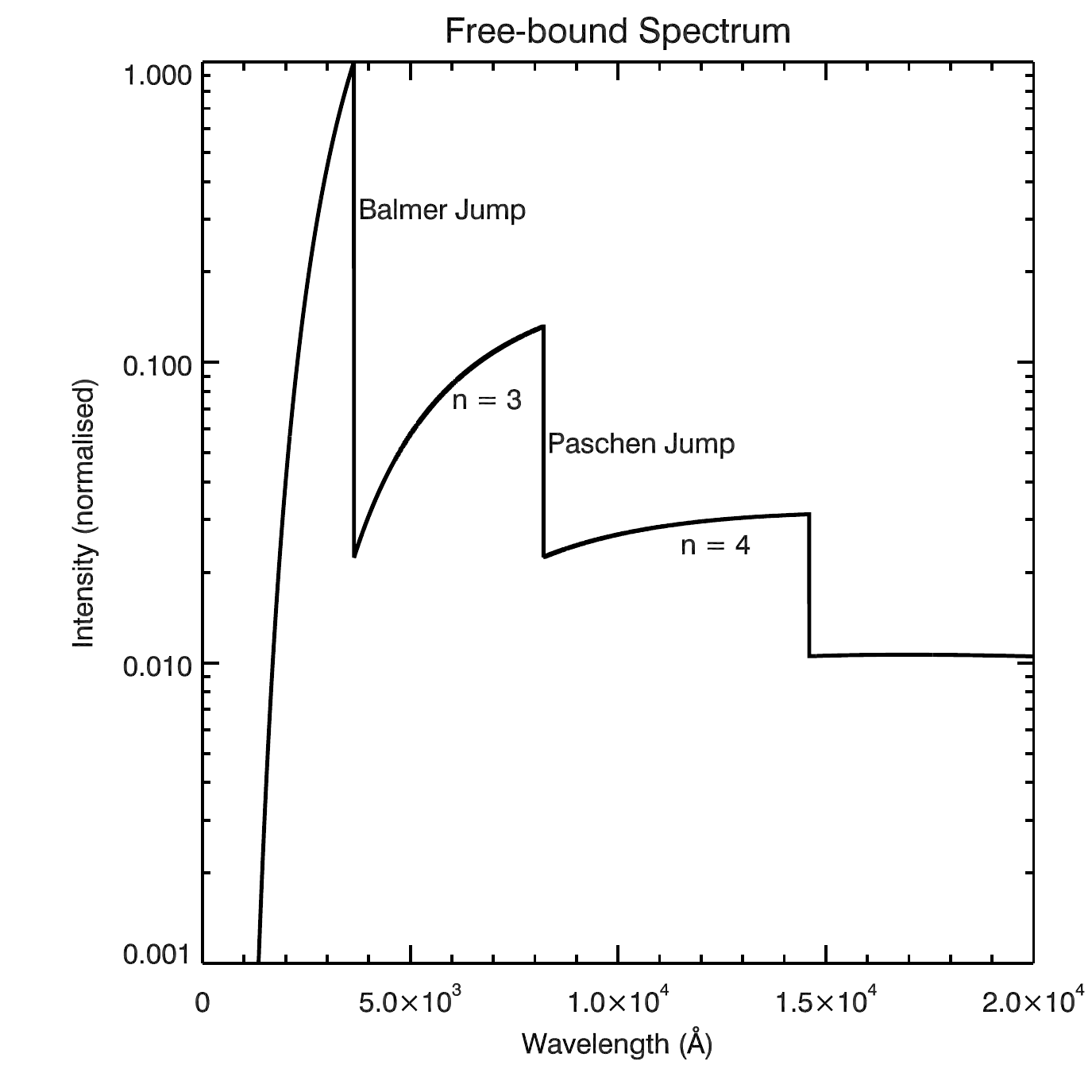}
\caption{\textsl{Example of a Hydrogen free-bound spectrum.}}
\label{fig:fbspectra}
\end{figure}

For each value of temperature within the overlap region in Figure~\ref{fig:val_err_left}, the upper and lower estimates of $\rm n_{e}^{2}$L$\rvert_{\rm \lambda,obs}$ (the value of $\rm n_{e}^{2}$L$\rvert_{\rm \lambda,obs}$ along the continuous white lines) were used in Equation~\ref{eq:Ifb} to produce model free-bound optically thin continuum spectra over a wavelength range $\lambda$~$\in$ [0-16000]~\AA. 
This was repeated for each time step. An example free-bound spectrum is shown in Figure~\ref{fig:fbspectra}. In this figure intensity has been normalized and temperature set to 8500~K. 

Integrating these spectra over wavelength at each time step yields the instantaneous energy flux,  
${F_{\rm fb}{\rm~ergs~s^{-1}~cm^{-2}~sr^{-1}}}$. 

\begin{figure}
			\centering
			\hbox{
				\subfloat[]{\label{fig:p_fb}\includegraphics[width=0.4\textwidth] {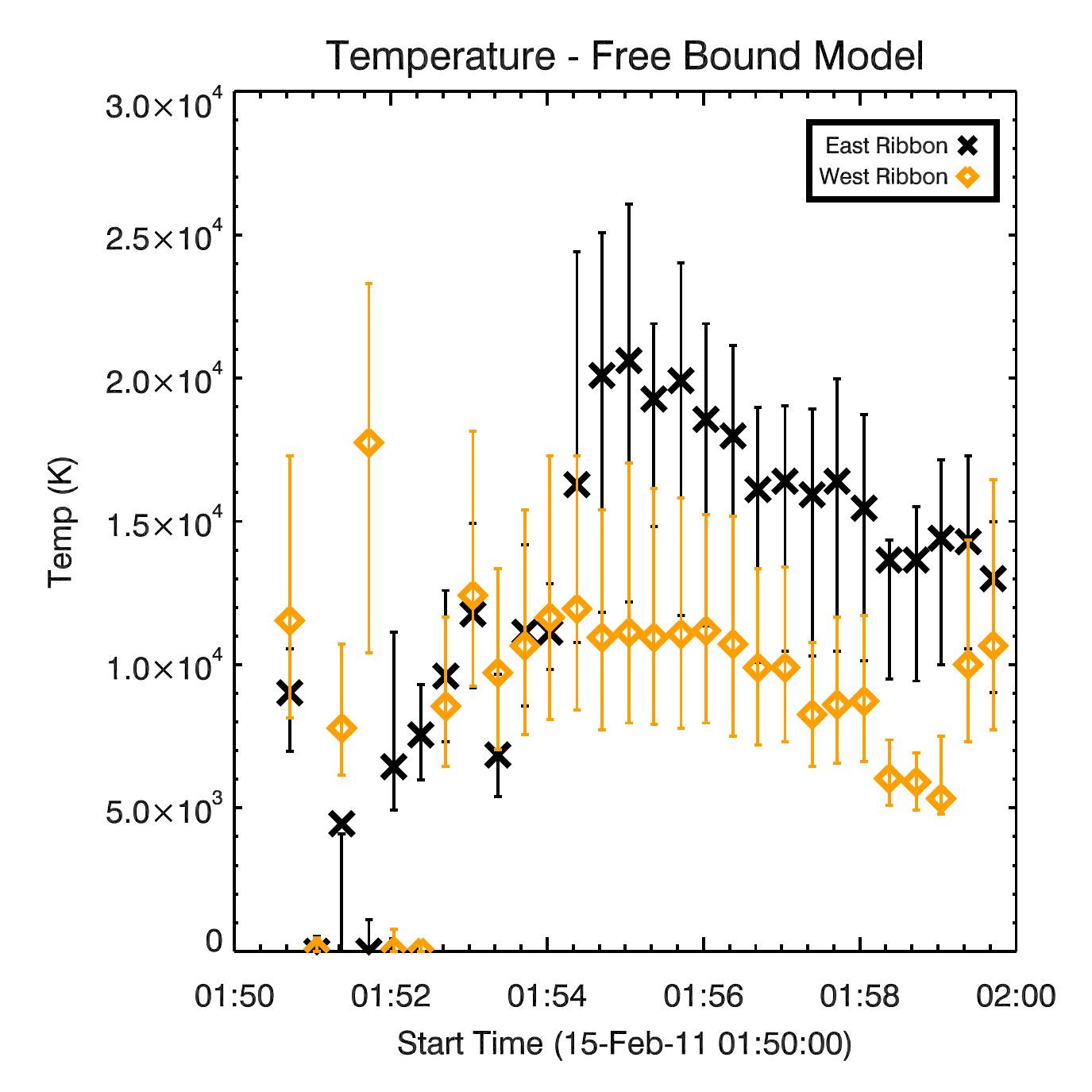}}
				}
			\hbox{
				\subfloat[]{\label{fig:pr}\includegraphics[width=0.4\textwidth]{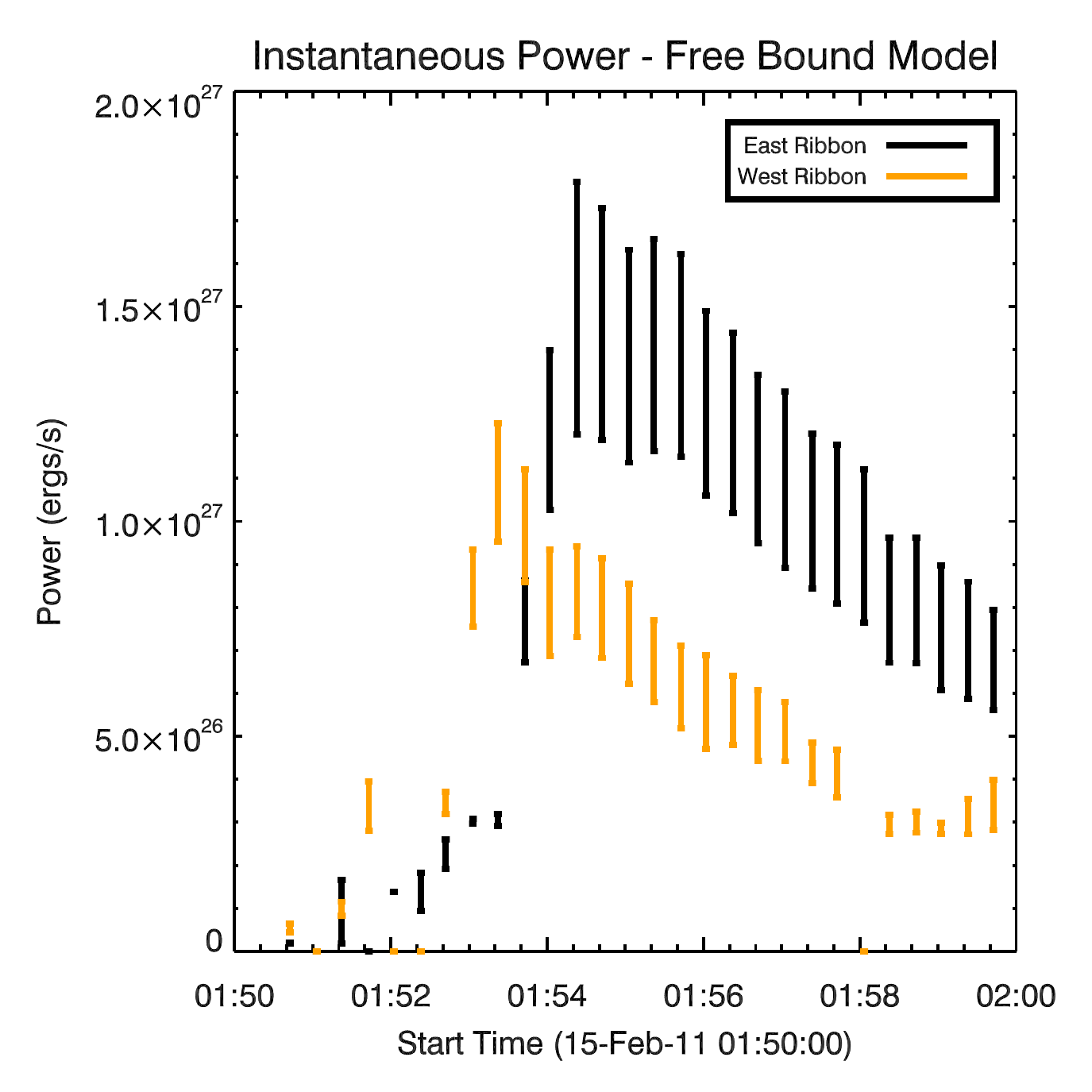}}  
				}
			\caption{\textsl{(a) The most probable temperature of WL sources, with ranges plotted as error bars, using the simplified free-bound model. (b) The upper and lower limits of instantaneous power from the free-bound model}}
		\label{fig:fb_evol}
		\end{figure}

In the optically thin model ($\tau \ll 1$) radiation emitted within the flaring slab is not attenuated, so the instantaneous power was calculated by multiplying the emitting area by F$_{fb}$, and by $\pi$ to account for emission in all directions.
\begin{equation}\label{eq:power_fb}
P_{fb} = \pi A F_{fb}
\end{equation}
$P_{\rm fb}$ was calculated at each time step, for each temperature value, and for the upper and lower limits, providing a range of possible power values since the actual temperature of the emitting slab was not known. This range is plotted in Figure~\ref{fig:fb_evol}.

The power was measured as being $\rm{P_{fb}}\sim10^{27}\rm{ergs~s}^{-1}$, somewhat higher than the values determined assuming a blackbody source.

Figure~\ref{fig:fb_evol}(a,b) shows the evolution of temperature and of instantaneous power of two sources we have been studying. The temperature plotted is the value for which the overlap of constraints from the red, green and blue channels was closest, but it should be remembered that there was a range of temperatures allowed as described above, which is represented as error bars in this plot. Power is presented as upper and lower limits since the density and slab thickness were not known. These plots exhibit the same general trends as observed previously, with the temperature in the western ribbon again being more ambiguous.

%%%%%%%%%%%%%%%%%%%%%%%%%%%%%%%%%%%%%%%%%%%%%%%%%%%%%
%%%%%%%%%%%%%%%%%%%%    DISCUSSION    %%%%%%%%%%%%%%%%%%%%%%%%
%%%%%%%%%%%%%%%%%%%%%%%%%%%%%%%%%%%%%%%%%%%%%%%%%%%%%

\section{Discussion and Conclusions}\label{sec:discuss}

We have analysed the event SOL 15 February 2011 using \textsl{Hinode} SOT/BFI continuum data and \textsl{RHESSI} HXR data. Isolating flare footpoints in SOT red, green and blue continuum filters, and using reference spectra to provide the absolute SOT calibration, as described in Section~\ref{sec:data_red}, we determine the flare intensity, optical luminosity and footpoint temperature under the assumption of (i) blackbody emission from the photosphere, and (ii) hydrogen recombination radiation from an optically-thin slab. We investigated one set of WL sources identified as first brightening in Frame 11 from the eastern ribbon, and one set of WL sources identified as first brightening in Frame 9 in the western ribbon, in order to show typical WLF characteristics in this event. 

Under assumption (i) we find temperature increases of approximately 200 K, whereas (ii) leads us to conclude a slab temperature in the range [5,500 - 25,000]~K, with $T \approx 20,000$~K at the peak . In the eastern ribbon color temperatures and effective temperature agree within errors, consistent with a blackbody interpretation (with the western ribbon showing some ambiguity). We have also calculated the flare optical power radiated under model assumptions (i) and (ii), finding an instantaneous power emitted by the newly brightened sources in the range $10^{26}$~ergs s$^{-1}$ for a blackbody source and $10^{27}$~ergs s$^{-1}$ for the free-bound slab. The observed radiated power was measured to be consistent with the blackbody model and the total energy emitted by WL over the course of the observations was found to be on the order of 10$^{(28-29)}$~ergs.

A photospheric origin for flare WL emission is difficult to explain, but is nonetheless implied by the consistency of color and effective temperatures in the eastern ribbon, and by considering the observed power output compared to model power output. Previous observations of a continuum enhancement near opacity minimum in the near infra-red \citep{2004ApJ...607L.131X} support a source in the lower atmosphere, and \cite{2013ApJ...776..123W} recently reported finding WL sources with a similar temperature increase as we observe here, suggesting a photospheric origin (though with with the different SOT channels separated in height). \cite{1982SoPh...80..113H} proposed an upper photospheric origin for flare optical emission, following observations of a temperature increase of a few hundred kelvin (consistent with our observations), and temperature-minimum heating is also implied by analysis of spectral line profiles \citep{1990ApJ...350..463M,1990ApJ...365..391M}. \cite{2007ApJ...656.1187F} and \cite{2010ApJ...715..651W} demonstrated that  with an electron beam cutoff around 20-30~keV, there is sufficient power in electrons to explain the flare WL emission. But the power-law spectrum means that the beam energy is concentrated around the cutoff, and such electrons certainly cannot reach the photosphere.

Our second model is an optically thin slab producing an enhanced free-bound continuum, as a result of heating or possibly direct collisional ionisation with beam electrons, but without allowing this continuum to produce backwarming of the photosphere. Our constraints suggest a density of 10$^{13-14}$~cm$^{-3}$, typical of mid-chromosphere. However, the temperature we find means this region must be heated. From the temperature constraints, total power and energy integrated across the free-bound continuum could also be estimated. The energy required is about an order of magnitude higher than those required by the optically thick model, but in terms of beam transport, a mid-chromosphere source is less challenging to explain than is a photospheric source. However, in other flares even this presents some problems \citep[e.g.][]{2007ASPC..368..423F}. An alternative source of mid-chomosphere heating may be by Alfv\'en wave damping as explored by \cite{1982SoPh...80...99E} and more recently by \cite{2013ApJ...765...81R}, who found it to be a viable heating mechanism for the temperature minimum region. 

\begin{figure}
\centering
\includegraphics[width=0.5\textwidth]{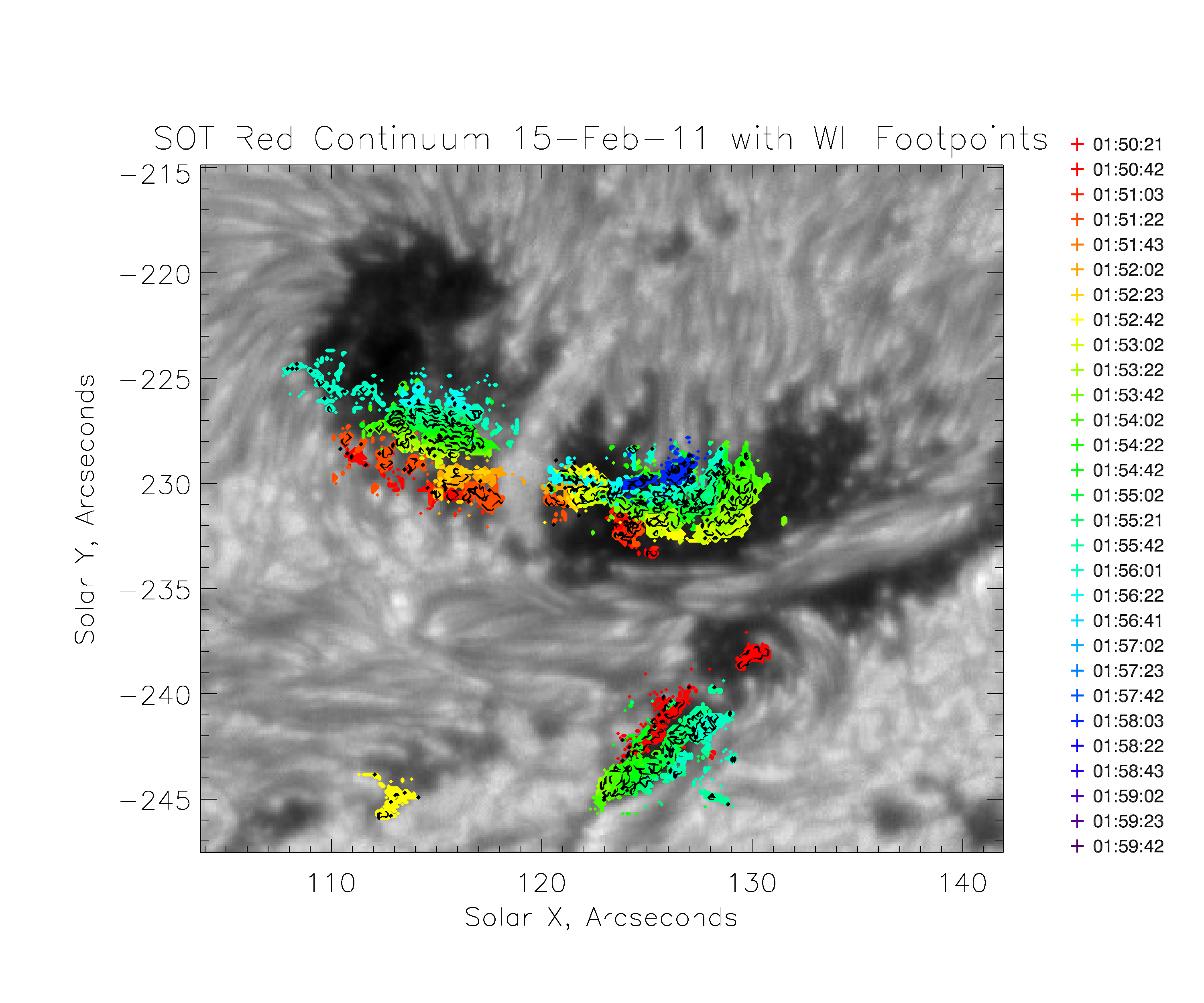}
\caption{\textsl{A comparison of the different thresholds - colored contours are 1-$\sigma$ data, and the black lines are contours drawn using 2-$\sigma$ data.}}
\label{fig:compare_sigmas}
\end{figure}

A model combining the radiation from a free-bound source and its radiative backwarming of the photosphere is a possibility that we have not considered here. \cite{2007PASJ...59S.807I} suggests that a `core and halo' structure in footpoints, as observed also by  \cite{2006ApJ...641.1210X} and \cite{1993ApJ...406..306N}, is consistent with such a model. Inspecting our flare pixels identified using thresholds 1-,2-, and 3- $\sigma$ above non-flaring levels shows that the brightest WL footpoint locations are in the middle of the larger area 1$\sigma$ results, illustrated in Figure~\ref{fig:compare_sigmas}. The implication could be that the main energy deposition is into a small central area of each flare footpoint.

The flare peak duration is around one minute, and the flare excess decays over the 5-10 minutes following the flare peak for which the optical data are available. 
Similar time profiles were observed by \cite{2006ApJ...641.1210X} who find a typical cooling time of 30~seconds (followed by a second stage lasting several minutes) in their 2~second cadence data and around 2~minutes (again, followed by a longer second stage) in their 1~minute cadence data.  Our timings seem largely consistent. \cite{2006ApJ...641.1210X} note that the decay timescales are similar to models of photospheric cooling, which would seem to favour the model of a photospheric blackbody enhancement.

It is essential that we obtain a deeper understanding of white light flares, in particular that we conclusively identify the primary emission mechanism of the energetically dominant optical and UV continuum radiation. WL enhancements are not simply a `Big Flare Syndrome' and are not unusual events, and we need to understand the low-energy continuum emission to fully appreciate flare energy transport mechanisms. This will require both imaging and spectroscopic observations of chromospheric flare emission, and numerical modelling of the radiative transfer.

\textsc{\\Acknowledgments:} \small{}
The authors would like to thank the anonymous referee for comments that have improved the clarity and quality of the paper, Dr. Ted Tarbell for calculating the Hinode/SOT response for us, and Drs. Hugh Hudson and Petr Heinzel for useful discussions. This research was improved following discussions between participants at an International Space Science Institute (ISSI) team meeting on Chromospheric Flares held at  ISSI in Bern, Switzerland. GSK acknowledges financial support of a postgraduate research scholarship from the College of Science and Engineering at the University of Glasgow. LF acknowledges financial support by the European Commission through HESPE (FP7-SPACE-2010-263086) and from STFC grant ST/l001808. \textsl{Hinode} is a Japanese mission developed and launched by ISAS/JAXA, collaborating with NAOJ as a domestic partner, NASA and STFC (UK) as international partners. Scientific operation of the \textsl{Hinode} mission is conducted by the \textsl{Hinode} science team organized at ISAS/JAXA. We are grateful to the data team for the open data policy.

\bibliographystyle{aa}
\bibliography{Kerr_Fletcher_WLF_SOT_ArXiV}

\end{document}